\documentclass[twocolumn]{aastex62}

\graphicspath{{./}{}}

\usepackage{amsmath}
\usepackage{amssymb}
\usepackage{soul}

\submitjournal{ApJ}

\shorttitle{Planetesimal Formation in Turbulence}
\shortauthors{Gole et al.}

\begin{document}

\title{Turbulence Regulates the Rate of Planetesimal Formation via Gravitational Collapse}
\correspondingauthor{Jacob B. Simon}
\email{jbsimon.astro@gmail.com}

\author[0000-0001-7369-8882]{Daniel A. Gole}
\affiliation{JILA, University of Colorado and NIST, 440 UCB, Boulder, CO 80309, USA}
\affiliation{Department of Astrophysical and Planetary Sciences, University of Colorado, Boulder, CO 80309, USA}

\author[0000-0002-3771-8054]{Jacob B. Simon}
\affiliation{Department of Physics and Astronomy, Iowa State University, Ames, IA, 50010, USA}
\affiliation{JILA, University of Colorado and NIST, 440 UCB, Boulder, CO 80309, USA}
\affiliation{Department of Space Studies, Southwest Research Institute, Boulder, CO 80302, USA}

\author[0000-0001-9222-4367]{Rixin Li}
\affiliation{Department of Astronomy and Steward Observatory, University of Arizona, 933 North Cherry Avenue, Tucson, AZ 85721, USA}

\author[0000-0002-3644-8726]{Andrew N. Youdin}
\affiliation{Department of Astronomy and Steward Observatory, University of Arizona, 933 North Cherry Avenue, Tucson, AZ 85721, USA}

\author[0000-0001-5032-1396]{Philip J. Armitage}
\affiliation{Center for Computational Astrophysics, Flatiron Institute, NY 10010, USA}
\affiliation{Department of Physics and Astronomy, Stony Brook University, Stony Brook, NY 11794, USA}

\begin{abstract}
We study how the interaction between the streaming instability and intrinsic gas-phase turbulence affects planetesimal formation via gravitational collapse in protoplanetary disks. Turbulence impedes the formation of particle clumps by acting as an effective turbulent diffusivity, but it can also promote planetesimal formation by concentrating solids, for example in zonal flows. We quantify the effect of turbulent diffusivity using numerical simulations of the streaming instability in small local domains, forced with velocity perturbations that establish approximately Kolmogorov-like turbulence. We find that planetesimal formation is suppressed by turbulence once velocity fluctuations exceed $\langle \delta v^2 \rangle \simeq 10^{-3.5} - 10^{-3} c_s^2$. Turbulence whose strength is just below the threshold reduces the rate at which solids are bound into clumps. Our results suggest that the well-established turbulent thickening of the mid-plane solid layer is the primary mechanism by which turbulence influences planetesimal formation and that planetesimal formation requires a mid-plane solid-to-gas ratio $\epsilon \gtrsim 0.5$.  We also quantify the initial planetesimal mass function using a new clump-tracking method to determine each planetesimal mass shortly after collapse.  For models in which planetesimals form, we show that the mass function is well-described by a broken power law, whose parameters are robust to the inclusion and strength of imposed turbulence. Turbulence in protoplanetary disks is likely to substantially exceed the threshold for planetesimal formation at radii where temperatures $T \gtrsim 10^3 \ {\rm K}$ lead to thermal ionization. Planetesimal formation may therefore be unviable in the inner disk out to 2-3 times the dust sublimation radius.
\end{abstract}

\keywords{Planet formation -- planetesimals -- protoplanetary disks -- stellar accretion disks}


\section{Introduction} 
Gravitational collapse models for planetesimal formation in protoplanetary disks posit the formation of solid particle clumps that exceed the Roche density, leading to the prompt formation of objects of km-scale and above. Models in this class are theoretically attractive because they work even if material barriers preclude collisional growth of particles above some size \citep{chiang10, johansen14}. They are also relatively robust to radial drift. Observational constraints are limited, but the mutual inclination distribution of primordial Kuiper belt binaries, for example, matches theoretical predictions for gravitational collapse initiated by the streaming instability \citep{nesvorny19}.  The matching colors and similar sizes of these binary components also agree with gravitational collapse models \citep{nesvorny10}.

Although there is a good general understanding of how planetesimals form via gravitational collapse, how the conditions for collapse are attained in protoplanetary disks remain unclear. The problem is the Roche density, which, measured against a baseline of a well-mixed disk with a solid-to-gas ratio $Z=0.01$, exceeds the initial particle density by a large factor of the order of $10^4$.  The most economical hypothesis is that a combination of vertical settling \citep[the key ingredient of first-generation gravitational collapse models;][]{goldreich73} and the streaming instability \citep{youdin05} suffice to concentrate solids to sufficiently high densities. Numerical simulations, however, show that even under favorable conditions (a mono-disperse particle distribution, with dimensionless stopping time $\tau \sim 0.1$) this minimal set of physical processes works only for super-Solar solid-to-gas ratios $Z \gtrsim 0.02$ \citep{johansen09,carrera15,yang17}. Less favorable conditions, for example because particles have $\tau \ll 0.1$ in the inner disk, or because streaming growth rates are suppressed due to a distribution in particle sizes \citep{krapp19}, exacerbate the problem. These considerations motivate models in which 
planetesimal formation is the culmination of a {\em multi-scale} process of particle concentration, involving vertical settling, and large-scale concentration through radial drift \citep{youdin02, youdin04}, photoevaporation \citep{throop05,alexander07,carrera17}, zonal flows \citep{johansen09b}, and vortices \citep{barge05,raettig15}. These processes create a favorable background on top of which small-scale concentration via the streaming instability, secular gravitational instability \citep{youdin11,takahashi14}, and a resonant drag instability, which adds vertical settling to streaming \citep{squire18}, can occur.

Turbulence may impose additional limits on planetesimal formation. The streaming instability is itself a source of turbulence in the gas, but here we are concerned with ``intrinsic" disk turbulence sourced from processes that do not depend directly on the presence of solid particles \citep[e.g. the non-ideal magnetorotational instability, or various hydrodynamic instabilities;][]{armitage11,turner14,gole18,lyra19}. Intrinsic gas turbulence diffuses coupled particles in all directions \citep{youdin07b}. This effect impacts the streaming instability by thickening the mid-plane particle layer \citep{dubrulle95}, leading to a reduced solid-to-gas ratio and slower linear growth rates. Turbulence can also suppress the streaming instability  \citep{youdin05}. \citet{umurhan19} formalized this suppression in the context of an $\alpha$ model \citep{shakura73} for isotropic disk turbulence. They found substantial modifications to the linear properties of the streaming instability in the presence of even weak turbulence, with $\alpha_{\rm turb} \sim 10^{-4}$. This can be compared to observational upper limits to the strength of turbulence in the outer regions of protoplanetary disks, which in different disks are at the level of $\alpha_{\rm turb} \sim 10^{-3}$ to $10^{-2}$ \citep{flaherty17,flaherty18}. Low levels of intrinsic turbulence could therefore, according to linear considerations of the streaming instability, severely impact planetesimal formation.

The influence of intrinsic turbulence on the non-linear evolution of the streaming instability\footnote{We follow convention and refer to simulations of aerodynamically-coupled  mixtures of particles and gas as ``streaming instability simulations", even though many such simulations include ingredients (such as vertical gravity, particle self-gravity, and gas compressibility) absent from the fiducial linear analysis.}, and on the rate of planetesimal formation via gravitational collapse, is relatively unexplored. \citet{johansen07} and \citet{johansen11} simulated particle-gas dynamics leading to gravitational collapse in a disk including turbulence driven by the magnetorotational instability \citep[MRI;][]{balbus98} in ideal magnetohydrodynamics. Planetesimal formation occurred in their models despite the presence of relatively strong turbulence, with an inferred $\alpha_{\rm turb} \simeq 0.001-0.003$. The apparent discrepancy between these results and the linear findings discussed above may be due to the fact that the MRI, in ideal MHD, generates local pressure maxima that passively (i.e. independent of particle-gas momentum feedback) concentrate solids. More recently, \cite{yang18} examined particle clumping (they neglected self-gravity) in the presence of turbulence driven by the MRI in both the ideal MHD limit and within an Ohmic dead zone. Despite significant temporal variability, they found a general trend for weak clumping with ideal MHD and lower $Z$ and stronger clumping in dead zones with higher $Z$.

In this paper we revisit the influence of intrinsic disk turbulence on planetesimal formation using high resolution simulations within a significantly smaller computational domain. In contrast to the previous work outlined above \citep{johansen07,johansen11,yang18}, which resolved the physical driving scale of the turbulence, we simulate a {\em small} purely hydrodynamic shearing-box, into which turbulence is injected using a prescribed forcing term in the momentum equation. We express the results in terms of an equivalent $\alpha_{\rm turb}$ by measuring the saturated velocity fluctuations within the gas. This approach allows us to attain the same high resolution (2560 grid points per gas scale height) that has been used in simulations without intrinsic turbulence \citep{johansen15,abod18}. We should note that -- while that is the effective resolution -- our boxes are smaller than a scale height, using 512 zones per dimension, with one simulation using 2048 in the z direction (see ~\ref{sec_stirTurb}).  The disadvantage to this setup, of course, is that by imposing hydrodynamic forcing on a small box, we exclude the possibility that MHD effects might materially modify the properties of the turbulence, and limit the extent to which the spectrum is modified by particle loading. These inherent drawbacks of our scheme should be borne in mind.

The layout of the paper is as follows. In \S2 we describe our methods, with a focus on the novel aspects developed for this study (the turbulent forcing, and alternate definitions of the initial planetesimal mass function). The results are presented in \S3. We find that the {\em rate} of planetesimal formation via gravitational collapse is a function of the strength of turbulence, and that there is no planetesimal formation for turbulent strengths at the higher end of the astrophysically interesting regime. When planetesimals do form, however, their initial mass function is not strongly dependent on the turbulent forcing. We discuss our results and conclude in \S4.  

\section{Methods}
In brief, we inject turbulence of specified strength by randomly forcing a small (generally $0.2 H$ on a side, where $H$ is the gas scale height) stratified shearing box containing an isothermal, un-magnetized fluid, and a solid component that is represented by super-particles. The gas and solid components are coupled by aerodynamic forces. Particle self-gravity is turned on after the combination of the streaming instability and the forced turbulence reach approximate saturation ($20 \Omega^{-1}$, where $\Omega$ is the orbital frequency, in our simulations). We then use a clump-finding algorithm to track the formation and evolution of gravitationally bound structures. The key diagnostics are the rate at which gravitationally bound clumps (``planetesimals") form, and their mass function evaluated shortly after the initial collapse.

Our methods for simulating the streaming instability and identifying planetesimals follow those described previously  \citep{simon16,simon17_streaming,abod18,li19}. We repeat only a high-level description of the common methods below, and refer readers to the earlier papers for full details of the implementation. The scheme for turbulent forcing, and the way we define the initial planetesimal mass function, are new and are described in detail.

\subsection{Streaming instability model}
\label{sec_SImodel}
We simulate the aerodynamic interaction between gas and solid particles in a local shearing-box domain \citep{hawley95}. For a flat, circular disk, in cylindrical polar co-ordinates $(R,\phi,z)$, we define a small Cartesian patch centered at $R=R_0$, which rotates at angular frequency $\Omega$. The local co-ordinates are defined via,
\begin{eqnarray}
 x & = & R - R_0, \\
 y & = & R_0 \phi
\end{eqnarray}
For an isothermal fluid with density $\rho$, velocity ${\bf u}$ and sound speed $c_s$, the hydrodynamic equations in the shearing box frame are,
\begin{eqnarray}
 \frac{\partial \rho}{\partial t} + \nabla \cdot (\rho {\bf u}) &=& 0, \\
 \frac{\partial \rho {\bf u}}{\partial t} + \nabla \cdot (\rho {\bf u}{\bf u} + P {\bf I}) &=& 
 2 q \rho \Omega^2 {\bf x} - \rho \Omega^2 {\bf z} \nonumber \\
 && - 2 {\bf \Omega} \times \rho {\bf u} + \rho_{\rm p} \frac{{\bf v}-{\bf u}}{t_{\rm stop}}, \\
 P &=& \rho c_s^2.
\end{eqnarray}
Here $\bf I$ is the identity matrix, and the terms on the right hand side of the momentum equation involve the shear parameter $q=3/2$, and a source term due to the aerodynamic coupling to the solids, which have density $\rho_{\rm p}$, velocity ${\bf v}$, and stopping time $t_{\rm stop}$. $P$ is the gas pressure. For those simulations that include forced turbulence, an additional source term (described below) is added to the momentum equation.

It is numerically advantageous to solve both the fluid and particle equations in a frame from which the background shear has been subtracted \citep{masset00}. Denoting the frame used for computations with a prime, for the gas velocity the transformation is,
\begin{equation}
 {\bf u}^\prime = {\bf u} - (q \Omega x) \hat{\bf y},
\end{equation}
where $\hat{\bf y}$ is a unit vector. The solids are represented by an ensemble of particles, $i = 1, 2, \ldots$, which in the primed frame follow the equation of motion,
\begin{eqnarray}
 \frac{{\rm d} {\bf v}_i^\prime}{{\rm d}t} =
 2 ( v_{iy}^\prime - \eta v_{\rm K} ) \Omega \hat{\bf x} - 
 (2-q) v_{ix}^\prime \Omega \hat{\bf y} - 
 \Omega^2 z \hat{\bf z} \nonumber \\
 - \frac{ {\bf v}_i^\prime - {\bf u}^\prime}{t_{\rm stop}} + {\bf F}_{\rm g}.
\end{eqnarray}
Here $\eta v_{\rm K}$ measures the difference in azimuthal velocity in the physical system between the unperturbed azimuthal gas velocity and the Keplerian velocity. ${\rm \bf F}_{\rm g}$ is the force per unit mass due to the self-gravity of the solids. When turned on, it is calculated as the gradient of a potential computed via Poisson's equation,
\begin{eqnarray}
 {\bf F}_{\rm g} &=& -\nabla \Phi_{\rm p} \\
 \nabla^2 \Phi_{\rm p} &=& 4 \pi G \rho_{\rm p}.
\end{eqnarray}
The solids are assumed to be collisionless, and the effect of self-gravity on the gas is ignored.

The boundary conditions are periodic in $y$ and shearing-periodic in $x$ for all variables (fluid, particle, and gravitational potential). In $z$, we use the outflow boundary conditions for the fluid quantities described by \citet{simon11a} and \citet{li18}, together with vacuum boundary conditions for the gravitational potential as described in the appendix of \cite{koyama09}.

The properties of the gas and particle system can be fully specified via several interchangeable sets of parameters. In physical units, the gas disk at a location where the angular frequency is $\Omega$ can be described by its mid-plane density $\rho_0$, sound speed $c_s$, and departure from Keplerian velocity $\eta v_{\rm K}$ (in turn determined by the radial gradient of density and temperature). The vertical profile of an isothermal disk, with scale height $H = c_s / \Omega$, is then,
\begin{equation}
 \rho(z) = \rho_0 \exp \left( -\frac{z^2}{2 H^2} \right), 
\label{eq_hydrostatic} 
\end{equation}
and the relation between the mid-plane density $\rho_0$ and the column density $\Sigma$ is,
\begin{equation}
 \rho_0 = \frac{1}{\sqrt{2 \pi}} \frac{\Sigma}{H}.
\end{equation}
From these physical quantities we can construct two dimensionless parameters,
\begin{eqnarray}
 \Pi & \equiv & \frac{\eta v_{\rm K}}{c_s}, \\
 \tilde{G} & \equiv & \frac{4 \pi G \rho_0}{\Omega^2},
\end{eqnarray}
which measure the importance of the headwind for solids and the relative strength of self-gravity and tidal shear, respectively.  The latter is directly related to the Toomre parameter for the gas disk, $Q = \sqrt{8/\pi}\tilde{G}^{-1}$. Characterizing the solids requires two more parameters, the ratio of solid to gas surface density,
\begin{equation}
 Z \equiv \frac{\Sigma_{\rm p}}{\Sigma},    
\end{equation}
and the dimensionless stopping time,
\begin{equation}
 \tau \equiv \Omega t_{\rm stop}.
\end{equation}
In the Epstein regime of drag $\tau$ maps directly to particle size \citep[e.g][]{armitage10, youdin10}. For spherical particles of radius $s$ and material density $\rho_{\rm m}$,
\begin{equation}
 \tau = \frac{2}{\pi} \frac{\rho_{\rm m}}{\Sigma} s.
\end{equation}
In this work we consider the case where the solids have a single $\tau$. 

The setup described above (or a close variant of it) has been used in many numerical simulations of the streaming instability. (An exception is the work of \citet{benitezllambay19}, who used a fluid rather than a particle description of the solids.) The model numerical system differs in several respects from the model analytic one studied by \citet{youdin05}. We include the vertical component of stellar gravity, model the gas as a compressible rather than an incompressible fluid, and treat the particles as a collisionless system rather than a pressure-less fluid.

\subsection{{\sc Athena} simulation methods}
We solve the hydrodynamic and particle equations simultaneously using the {\sc {\sc Athena}} code \citep{stone08}, in a shearing box geometry \citep{stone10} with orbital advection \citep{masset00}. The {\sc {\sc Athena}} configuration used to solve the fluid equations uses the unsplit Corner Transport Upwind (CTU) integrator, third-order (piecewise parabolic) reconstruction, and an HLLC Riemann solver. The coupling to the particle dynamics requires additional choices, most importantly in how to interpolate between grid-based and particle-based quantities (e.g. how to define ${\bf v}$ from a set of particles with ${\bf v}_i$), in how to integrate the particle trajectories, and in how to compute ${\bf F}_{\rm g}$. For the non-self-gravitating dynamics we use the particle implementation described by \citet{bai10}. For the particle self-gravity we use the FFT approach described by \citet{simon16}, which is an extension of the gas self-gravity scheme of \citet{koyama09}.

The gas is initialized in an equilibrium state with velocity given by the Keplerian shear and a vertical density profile $\rho = \rho(z)$ specified by Equation~(\ref{eq_hydrostatic}). For the particles, no initial equilibrium state exists for streaming instability simulations that include the vertical component of stellar gravity but omit intrinsic turbulence. While an equilibrium (between vertical settling and turbulent diffusion) can be found for a tracer population of particles in the presence of turbulence \citep{dubrulle95}, our simulations include control runs (with no turbulence), and turbulent models where the particle loading is sufficiently strong that simple equilibrium models would only be approximate. We have therefore chosen to initialize the particles in all of our runs in a thin layer, using a Gaussian vertical density distribution, with $H_{\rm p} = 0.025 H$. The in-plane velocities are set according to the \citet{nakagawa86} solution (which is an exact equilibrium for an unstratified disk).  

The use of outflow boundary conditions imposed at $z \ll H$ leads to an unphysical loss of gas from the simulation domain. We compensate for this by renormalizing the gas density at each time step to maintain the initial value of the gas mass.

A common set of parameters is adopted for all of the runs. We take $\rho_0 = 1$, $c_s = 1$, and $\Omega = 1$ (all in code units). For the dimensionless parameters we take $\tau = 0.3$, $Z = 0.02$, $\Pi = 0.05$ and $\tilde{G} = 0.1$. The use of these parameters places us in a region of parameter space that is relatively easy to study numerically, and allows for comparison with a number of previous simulations of the streaming instability. 

These parameters lead to a natural choice for a mass scale: the ``Gravitational Mass", given by
\begin{equation}
M_G = \frac{\sqrt{2}}{2} \pi^{9/2} Z^3 \widetilde{G}^2 \rho_0 H^3.
\end{equation}
Physically, this corresponds to the mass enclosed by a circle with surface density $\Sigma_p$ and a diameter equal to the critical Toomre wavelength for gravitational collapse:
\begin{equation}
\label{eq_toomre}
    \lambda_G = \frac{4 \pi^2 G \Sigma_p}{\Omega^2}
\end{equation}
Unless otherwise stated, all masses throughout the paper will be in units of $M_G$.  

\subsection{Turbulent forcing}
\label{sec_stirTurb}
The properties of turbulence in protoplanetary disks depend upon the nature of the driving, shear, magnetohydrodynamic effects, and particle-gas coupling. To reduce the problem to a one-parameter family of solutions, we assume that on the small scales where gravitational collapse occurs turbulence is adequately described by Kolmogorov scalings \citep{kolmogorov41}. In this limit, the power spectrum of fluid velocity fluctuations scales as a power law with the spatial wavenumber $k$,
\begin{equation}
E_1(k) \propto k^{-n}, \hspace{3mm} n=5/3,
\end{equation}
where $k$ is the magnitude of the wavenumber (i.e. $k=\sqrt{k_x^2+k_y^2+k_z^2}$ in 3D Cartesian coordinates) and $E_1(k)$ is a linear energy density in $k$-space such that $E_1(k) dk$ has units of energy.    

To maintain a turbulent fluid state within the simulation domain, we impose scale-dependent driving of the velocity field using a scheme similar to that described by \cite{dubinski95}. In brief, each mode in 3D $k$-space ($k_x, k_y, k_z$) is assigned a scale-dependent amplitude and a random phase between 0 and $2\pi$. The velocity perturbations in real space are then computed using an inverse Fourier transform.  These velocity perturbations are re-calculated and applied with a cadence of $10^{-3}\Omega^{-1}$. 
Appendix \ref{app_stirTurb} gives a full discussion of the implementation, together with the results of tests that confirm that our methods produce a box of homogeneous, isotropic turbulence with the intended power spectrum. 

The control parameter for the amplitude of the turbulent fluctuations is the total energy injection rate. To relate this to quantities more commonly used in disk studies, we calculate and report an effective $\alpha_{\text{drive}}$ parameter in the saturated state via,
\begin{equation}
\label{eq_alpha}
\alpha_{\text{drive}} c_s^2 = \sum_{i=x,y,z}^{} \big(\overline{\langle |\delta v_i(x,y,z,t)| \rangle}\big)^2,
\end{equation}
where $c_s$ is the sound speed and,
\begin{equation}
\delta v_i(x,y,z,t) \equiv v_i(x,y,z,t)-\langle v_i \rangle_{xy}(z,t).                  
\end{equation}
The brackets indicate a spatial average, with the $xy$ subscript denoting an average over the x-y plane. The over-bar indicates an average over time. In Kolmogorov turbulence there should be a linear relation between the energy injection rate and $\alpha_{\text{drive}}$. Due to the presence of open vertical boundaries, this is not exactly true in our simulations.  This effect is significant enough that it merits a correction to produce turbulence of the desired magnitude across a range of values for $\alpha$.  Appendix \ref{app_stirTurb}  describes how we empirically develop a mapping from the energy injection rate to the resulting $\alpha_{\text{drive}}$ value in the saturated state of turbulence.  To give a general feel for the degree of non-linearity: changing the turbulence level from $\alpha=10^{-5}$ to $\alpha=10^{-3}$ requires a $10^{2.35}$ increase in the energy injection rate -- in contrast to $10^2$ if the scaling was purely linear.  Note that throughout this paper we will be careful to refer to this driven turbulence with ``$\alpha_{\text{drive}}$", general turbulence with $\alpha_{\text{turb}}$, and will save ``$\alpha$" to be used as a power law exponent.  

It is important to note that $\alpha_{\text{drive}}$ as defined above is a dimensionless measure of the turbulent kinetic energy, that in principle is quite distinct from the Reynolds stress or the total angular momentum transport rate parameterized by the \citet{shakura73} ``$\alpha$". In fact, our randomly forced fluid turbulence generates only very small Reynolds stresses. We choose to define $\alpha_{\text{drive}}$ in terms of the kinetic energy to make contact with the notion of turbulent diffusion.

We perform 4 production-resolution simulations with varying driven turbulence strengths,
\begin{eqnarray}
 \alpha_{\rm drive} & = & 0 \,\,\, {\rm (``control")}, \nonumber \\
 \alpha_{\rm drive} & = & 10^{-4} \,\,\, {\rm (``weak \, turbulence")}, \nonumber \\
 \alpha_{\rm drive} & = & 10^{-3.5} \,\,\, {\rm (``moderate \, turbulence")}, \nonumber \\
 \alpha_{\rm drive} & = & 10^{-3} \,\,\, {\rm (``strong \, turbulence")}. \nonumber  
\end{eqnarray}
With the exception of the strong turbulence simulation, our production runs have $512^3$ gas zones and the same number of particles, in a cubical domain $0.2 H$ on a side.  For the strong turbulence simulation a taller box is required to ensure that particles are not escaping out the top and bottom of the box en masse.  In this case we extend the domain to be $0.8H$ in the vertical direction and scale the resolution accordingly, resulting in a $512 \times 512\times 2048$ gas zones and the same number of particles.  Our simulations are initialized with no particle self-gravity and allowed to come to a roughly steady state for 20 $\Omega^{-1}$.  At this point, self-gravity for the particles is turned on, allowing clumps of sufficient density to become gravitationally bound.  Unless otherwise noted, we define the time at which self-gravity is enabled to be $t=0$.

\subsection{Identification of Planetesimals}
\label{sec_planAlg}
To identify gravitationally bound clumps, we use the PLanetesimal ANalyzer code (PLAN for short, \cite{li19}).  PLAN uses the same approach as HOP \citep{eisenstein98}, a code used to identify dark matter halos in cosmological simulations.  First, particles are stored in a data structure based on a Barnes-Hut tree \citep{barnes86}, which allows for very fast searches of a given particle's neighbors.  Then, particles in dense regions are identified using the HOP algorithm, which is capable of grouping together physically related particles.  For particles whose immediate surroundings have a particle density greater than a customized threshold of $\delta_{\text{outer}}=(8/9)\rho_r$, where the roche density $\rho_r$ is defined to be 
\begin{equation}
\frac{\rho_r}{\rho_0} = \frac{9}{\widetilde{G}},
\end{equation}
the algorithm looks for each particle's densest neighbor.  This process is repeated, creating a chain of particles with each in a denser region than the last, until a true peak is reached.  All particles that get mapped to the same peak are now tentatively considered a group.  PLAN then draws the boundaries between groups to find true, gravitationally bound clumps.  Intersecting groups are merged into a single clump if they are bound based on a calculation of their total kinetic and gravitational energy.  However, they are not merged if the density in the saddle point between them drops below $2.5\delta_{\text{outer}}$.  PLAN then discards any clumps with a Hill radius smaller than a single grid cell or with a peak density less than $3\delta_{\text{outer}}$.  Finally, PLAN outputs a list of the gravitationally bound clumps and their properties as calculated based on each clump's member particles.        

In considering the best way to measure planetesimal masses, we must account for a limitation common to fixed grid simulations: clumps will not collapse further than approximately the grid-scale because that is the scale at which the gravitational force is discretized (see \citealt{simon16}). Even at a resolution of $512^3$ in a box that is $0.2H$ on a side, the grid-scale is $\sim4\times10^{-4}H$.  At a radius of 10AU in a disk with an aspect ratio of $H/r=0.05$, this corresponds to a size of $\sim 3 \times 10^4$ km.  This scale is about 5 Earth radii, and is of the order of 1000 times larger than a typical planetesimal ($\sim 1$ to $100$ km).  Clumps in our simulations will have an enhanced size compared to the actual scales of planetesimals.  Consequently, we do not expect our simulations to correctly capture the post-formation evolution of planetesimals (e.g., mergers between planetesimals will not, in general, be accurately treated). To get closer to measuring the initial mass function we have developed a novel approach, described in the following section, for measuring their masses. In short, we track clumps from one frame to the next and create a mass-history for each clump.  From this data, we can then look at every clump that ever existed in the simulation and measure its mass {\it at the time of formation}.

\subsection{Clump Tracking Method}
\label{sec_trackingAlg}
\begin{figure}[h]
\centering
\includegraphics[width=0.99\columnwidth]{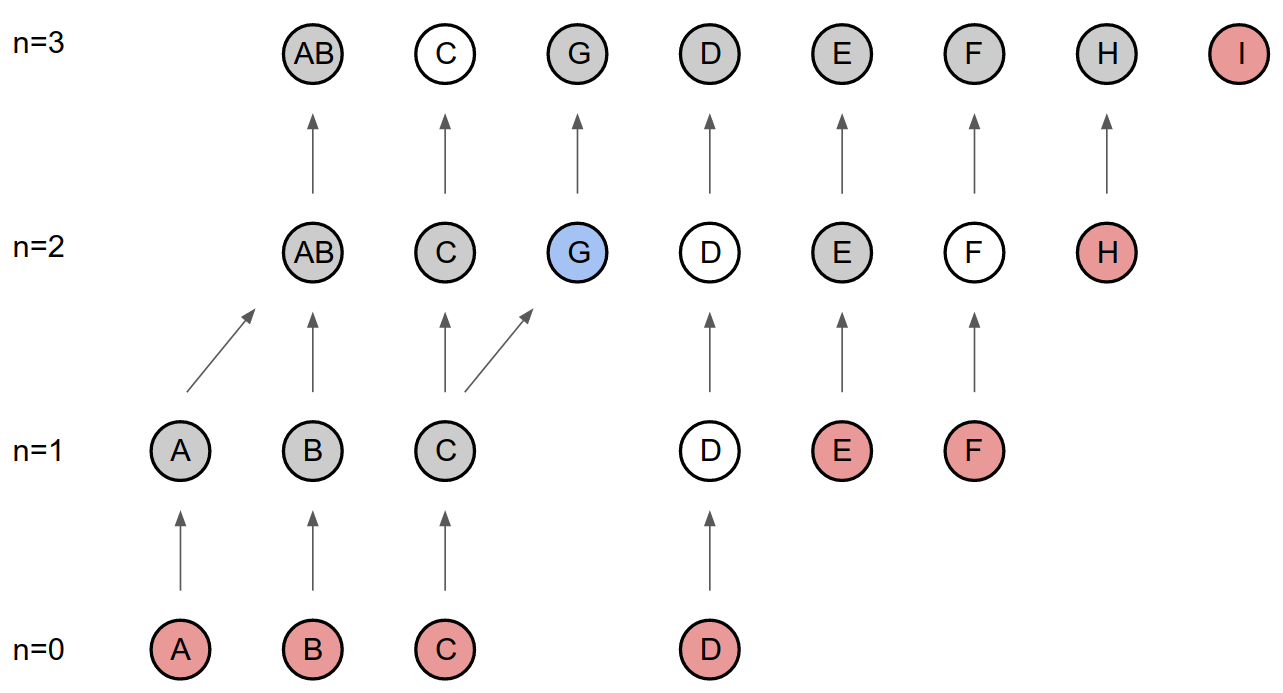}
\caption{An illustration of our particle tracking scheme.  The example shown here is intended to be demonstrative of our scheme and does not represent actual data from a simulation.  Each circle represents a gravitationally-bound clump.  Red corresponds to a new clump, gray to a previously found clump that was identified again, white to a clump that was not found (but will continue to be looked for in future frames), and blue to a new clump that formed from a split.  Clumps are consistently labeled from frame to frame.  Clumps that merge are given 2 labels and are continued to be tracked as both clumps.  In taking statistics, we use the masses of all of the red clumps.}
\label{fig_5_clumpTrackingScheme}
\end{figure}

PLAN can output lists of the particles belonging to each clump it finds, but does not use a persistent clump labeling scheme from frame to frame. We develop an algorithm to look through these particle lists and track clumps over time, illustrated graphically in Figure \ref{fig_5_clumpTrackingScheme}, and outlined as follows:
\begin{enumerate}
\item At the starting frame, the particle lists corresponding to each existing clump are read in and each clump is given an internal ID that will be consistent from frame to frame.
\item The particle lists for each clump in the following frame are read in.  For each clump in the current frame, the number of overlapping particles of each clump in the new frame is calculated.  This presents several possible outcomes: 
\begin{itemize}
    \item Case A: Only one new-frame clump has overlapping particles with the prior clump.  These clumps are identified as being the same clump and the clump's particle list is updated with the new-frame particles.  While not obvious, this case also covers merge scenarios -- two or more prior clumps can have a ``case A match" with the same new-frame clump.  When taking statistics, properties are determined by the initial particle list for each clump.    
    \item Case B: Multiple new-frame clumps have overlapping particles \footnote{When we discuss "overlapping particles" in this description we mean it in the temporal sense.  As an example: the particle list for clump A in frame 0 should be almost completely overlapping with the particle for clump A in frame 1 in most cases.  Within a given frame, different clumps will not have overlapping particles.} with the prior clump, indicating there has been a split of some sort.  In this case, the clump with more overlapping particles is marked as the same clump and its particle list with the new-frame particles.  If the other clump with overlapping particles is newly-formed this frame and more than 50\% \footnote{We find that the “split fraction” for newly formed clumps is close to either 1 or 0 in most cases.  In other words, when a new clump forms that has overlapping particles with a previous clump, the two most likely scenarios are that almost all of its particles came from an already existing clump, or very few (but not zero) particles came from an already existing clump.  Given this, the effect of this admittedly arbitrary choice for the split fraction threshold has a limited impact on the resulting mass distribution.} of its particles came from the parent clump, it is flagged as a ``splitter".  It will continue to be tracked from this frame on, but it will not be counted towards the statistical analysis.   
    \item Case C: No new-frame clumps have overlapping particles with the prior clump.  The clump has ``disappeared".  It could have been destroyed by turbulence or an interaction with other clumps, or it could have been very marginally gravitationally bound in the first place such that PLAN would identify it as gravitationally bound in one frame, but not the next.  This happens relatively frequently.  In this case, the clump is not matched with any new-frame clumps and its particle list is not updated.  However, the algorithm will continue to look for this clump in subsequent frames with the old particle list.  The initial properties of these clumps are still counted towards the final analysis. 
\end{itemize}
\item At this point clumps that existed in the previous frame have been mapped to the new frame.  Now, new clumps need to be dealt with (i.e. clumps that are ``unclaimed" by clumps that existed in the previous frame).  The particle lists for these new clumps are read in and their matches will be looked for in subsequent frames.
\item Continue to step through the frames by repeating steps 2 and 3.
\end{enumerate}

Measuring initial masses with this method is able to solve 3 problems stemming from the artificially large size of simulated clumps:
\begin{enumerate}
    \item Accretion. After formation, clumps will accrete particles at an artificially enhanced rate.  Examining only the initial masses ignores this accretion.
    \item Mergers. Clumps will merge at an enhanced rate.  If two clumps merge, a snapshot would count them as one large clump. In our tracking method, we measure the initial formation mass of the two separate clumps.
    \item Splits and Tidal Stripping. If a clump was able to fully collapse, the rate of clumps being tidally stripped and split into multiple clumps should be quite low.  However, this occurs quite often with the artificially large clumps in these simulations, which are less bound and more easily tidally stripped.  Our algorithm is able to ignore this effect by measuring only the initial mass of the parent clump.  
\end{enumerate}

\section{Results}

\begin{figure*}[t]
\includegraphics[width=1.1\columnwidth]{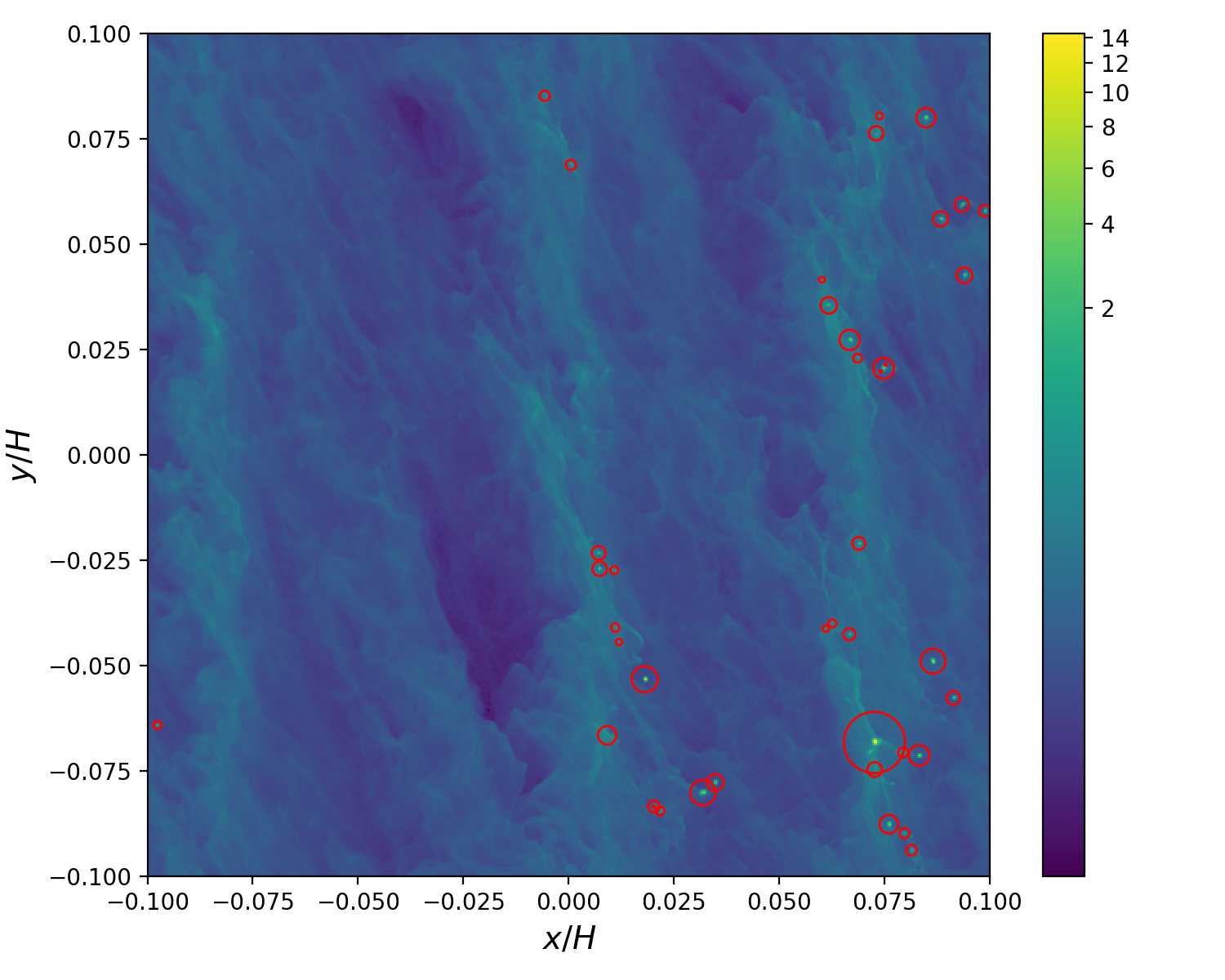}
\hspace{-5mm}
\includegraphics[width=1.1\columnwidth]{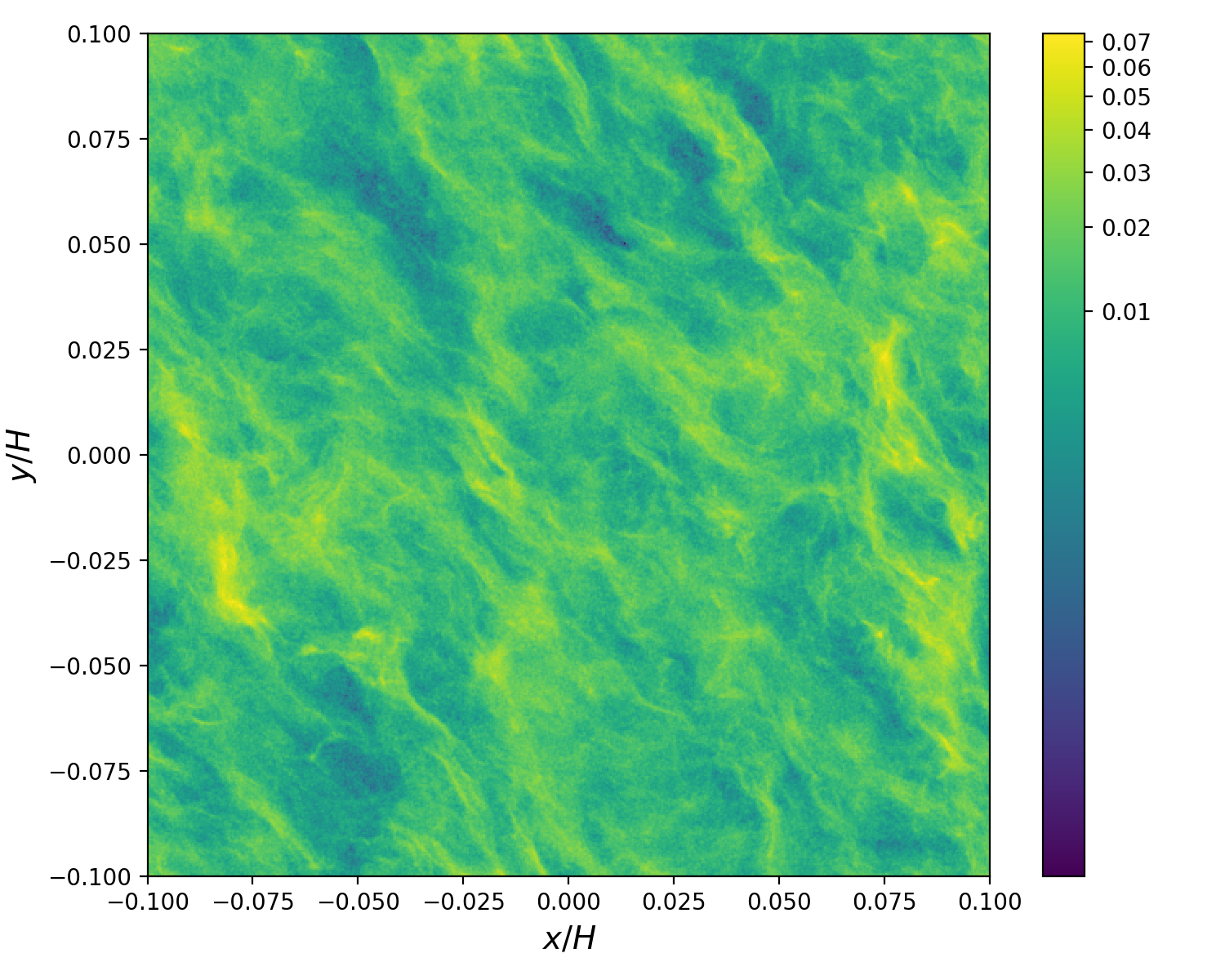}
\includegraphics[width=0.95\columnwidth]{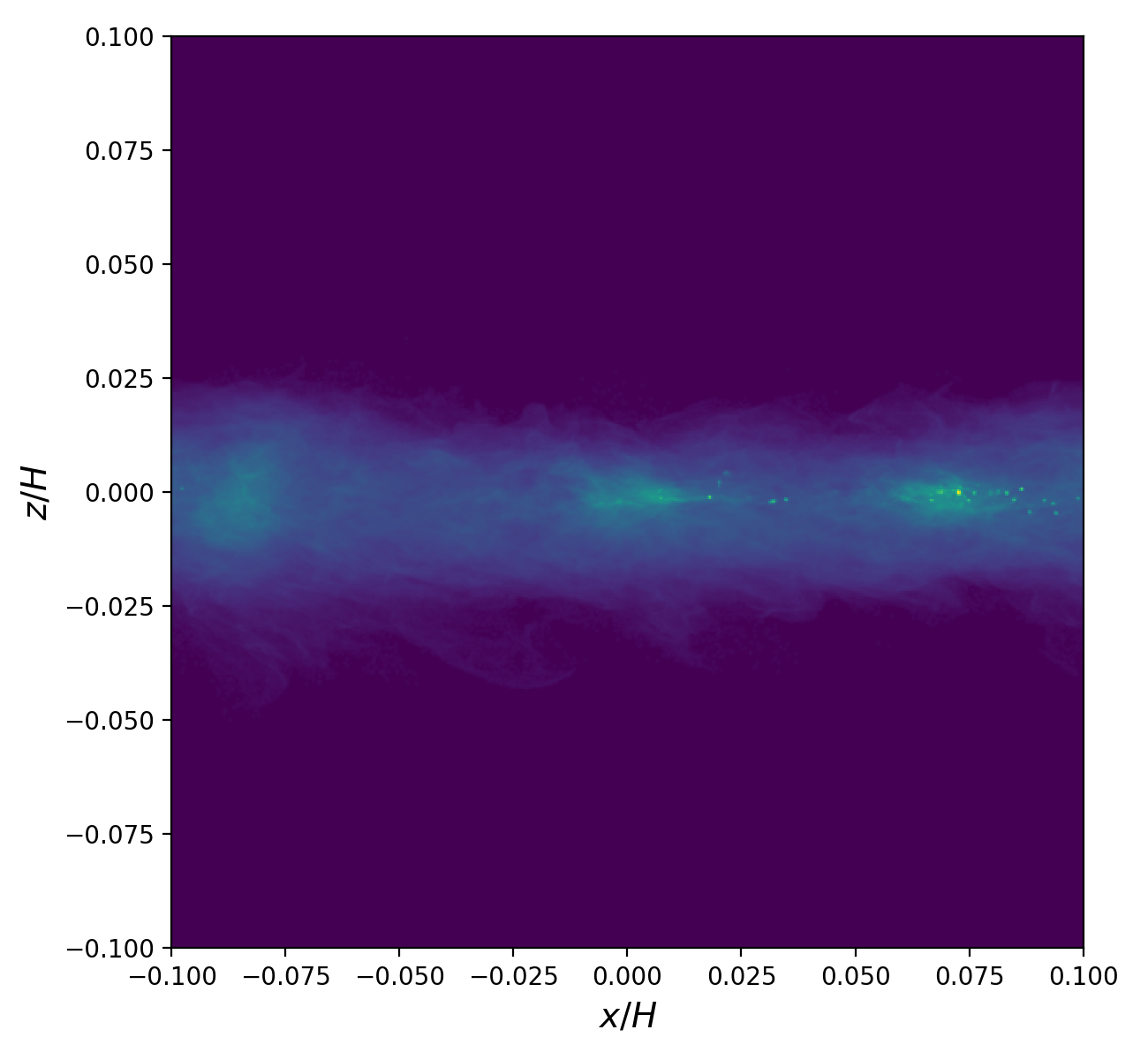}
\hspace{6.5mm}
\includegraphics[width=0.95\columnwidth]{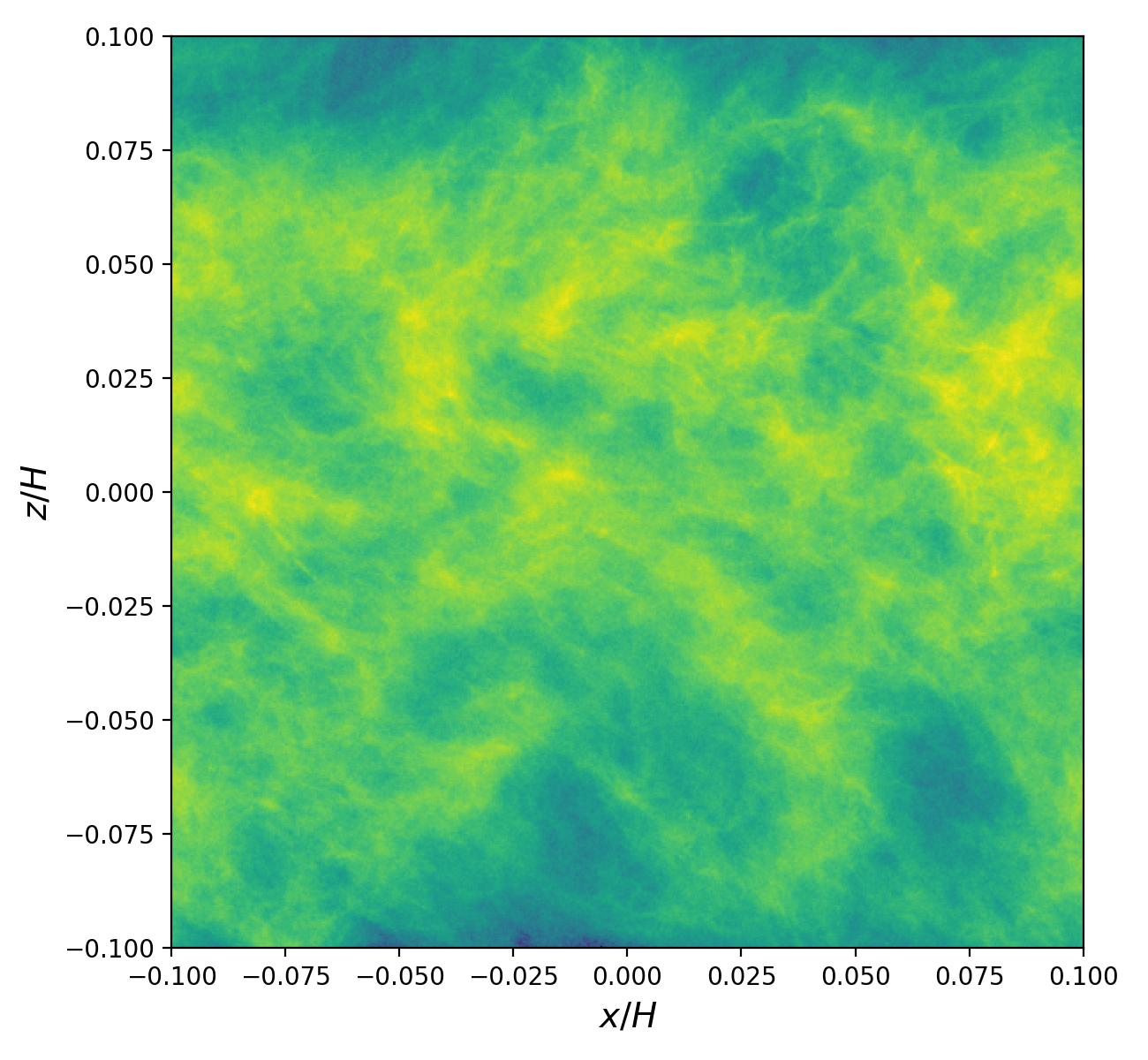}
\caption{Snapshots of the particle density, in units of the Roche density, for the weak turbulence simulation (left) and for the strong turbulence simulation (right).  The top row shows a narrow slice of the $x$--$y$ plane at the mid-plane, while the bottom plot shows the density in the $x$--$z$ plane, averaged over the $y$ direction. Note that in the strong turbulence case, the simulation domain extends between $\pm 0.4 H$, and is larger than the visualized portion.
Snapshots were taken 8 $\Omega^{-1}$ after self-gravity had been enabled.  The red circles in the weak turbulence plots show the clumps identified by PLAN, with the area of the circle proportional to the mass of the clump.  Note that the color-bars are different between the two runs.  While the strong turbulence run shows SI-like structure, it is unable to concentrate solids to near the Roche density, after which SG would take over and create bound clumps. The weak turbulence run does exhibit this collapse of particles into clumps, creating a difference in maximum particle density of several orders of magnitude despite only a single order of magnitude difference in $\alpha_{\text{drive}}$.} 
\label{snapshots}
\end{figure*}

The structure of the particle density field in the presence of driven turbulence is shown in Figure~\ref{snapshots}, which shows slices and projections of the density in the $x-y$ and $x-z$ planes from the weak and strong turbulence simulations. The most striking difference is the thickness of the particle layer, which is largely confined within $\pm 0.025 H$ in the weak turbulence run but which extends to about $\pm 0.075 H$ in the strong turbulence case (note that for this reason the strong turbulence run employed a taller box). The mid-plane slices both show filamentary structure in the solids, which is more axisymmetric in the case of weak turbulence. {\em Much} higher densities are attained in the weak turbulence run, leading to gravitational collapse and the prompt formation of bound clumps. No such clumps are observed in the presence of strong turbulence.

\subsection{The rate of planetesimal formation}

We find that planetesimals form in every simulation except one: there are no gravitationally-bound clumps in the simulation being driven at $\alpha_{\text{drive}}=10^{-3}$ after 18 $\Omega^{-1}$ with self-gravity on.   At slightly lower levels of $\alpha_{\text{drive}}=10^{-3.5}$, gravitationally bound clumps do form, meaning that the threshold for turbulence to disrupt planetesimal formation via the streaming-instability/self-gravity (SI/SG) mechanism is somewhere between $\alpha_{\text{drive}}=10^{-3.5}$ and $10^{-3}$.  Equivalently, the threshold in terms of $\delta v / c_s$ would be between 0.018 and 0.032. This result suggests that even moderately low levels of turbulence can preclude planetesimal formation. An immediate and obvious consequence is that we expect a fully MRI active disk (ionized throughout) to easily exceed this threshold.  However, at many locations within a disk, we expect non-ideal MHD effects to reduce the strength of the MRI at the mid-plane, creating a region of dampened turbulence\footnote{One exception to this is the Hall-dominated region, where turbulence can still persist via the Hall-shear instability \citep{kunz08}, though the degree to which this happens depends on the magnetic field orientation and the strength of Ohmic and ambipolar diffusion \citep{simon15b,bai15}}, permitting planetesimal formation. We save a full discussion of the consequences and implications of this threshold for \S\ref{sec_discuss}.  

\begin{figure}[h]
\centering
\includegraphics[width=0.99\columnwidth]{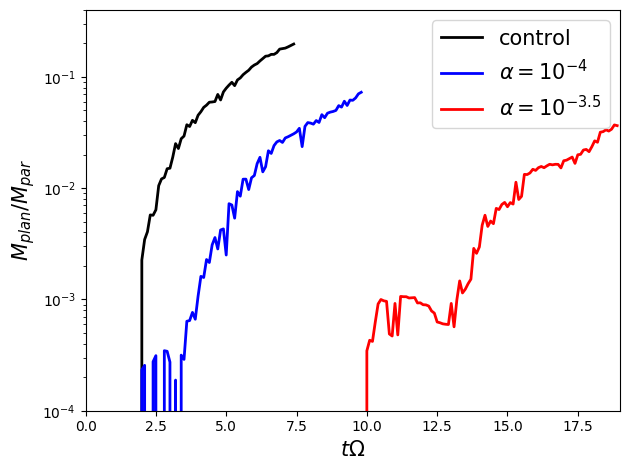}
\caption{The fraction of total particle mass contained in gravitationally bound clumps versus time for each of our simulations.  Note that $t=0$ is defined to be the time when self-gravity is turned on for the particles, which is after $20 \Omega$ in all of our simulations.  The difference in starting points on the plot is due to turbulence delaying clump formation. }
\label{fig_massFrac}
\end{figure}

To quantify the extent to which planetesimals form, a useful metric is the total amount of mass that ends up in planetesimals relative to the total mass of all particles in the simulations, $M_{\text{plan}} / M_{\text{tot}}$. Figure \ref{fig_massFrac} shows the time evolution of the planetesimal mass fraction over time for each simulation. Increasing the strength of forced turbulence both delays the onset of planetesimal formation, and reduces the average rate of planetesimal formation once it begins. Table \ref{tab_highLevel} shows a summary of the results for the number of clumps that form, and for the planetesimal formation rate in terms of this mass fraction. (We note that for simplicity of interpretation, we have not used our clump tracking algorithm here.) The rate at which solid mass is converted into bound clumps drops by approximately an order of magnitude between our control and moderate turbulence runs, before being cut-off entirely in the strong turbulence case. 

Clearly, turbulence reduces the amount of mass converted to planetesimals.  We will quantify the properties of the planetesimal mass distribution in the following section, but these result raise an immediate question: is the reduced mass in planetesimals a result of a smaller number of clumps forming, or instead the result of smaller {\it sized} clumps forming?  We find that the number of clumps measured in each simulation scales roughly in proportion to the formation rate, indicating that the average mass of the planetesimals that form is roughly constant in all of the simulations. The primary cause of the reduced mass-formation rate in the presence of turbulence is that there are fewer gravitationally bound clumps, not that the clumps formed are smaller.   

\begin{deluxetable}{l l c c c c}
\tablecaption{Planetesimal Formation Summary}
\tablewidth{0.99\columnwidth}
\tablehead{\colhead{Label} & \colhead{$\alpha_{\text{drive}}$} & \colhead{$t_{\text{end}}$} & \colhead{$N_{\text{clumps}}$} & \colhead{$M_{plan}/(M_{\rm tot} t_{\rm end}\Omega)$}} 
\startdata
Control     & 0.0         &   7.5   &  164   &  2.7\%   \\ 
Weak        & $10^{-4}$   &   9.8   &  76    &  0.74\%  \\
Moderate    & $10^{-3.5}$ &   19    &  29    &  0.19\%  \\
Strong      & $10^{-3}$   &   20    &  n/a   &  0\%    \\
\enddata
\label{tab_highLevel}
\tablecomments{A summary of clumping behaviour in our simulations.  $t_{\text{end}}$ denotes the time at which the simulation was ended.  The rightmost column indicates the average fractional formation rate: the amount of mass in gravitationally bound clumps relative to the total mass in particles, divided by the total amount of time with self-gravity on.}
\end{deluxetable}

\subsection{Initial planetesimal mass function}
\label{sec_IPMF}

To measure the initial mass function of gravitationally bound clumps we use the methods described in \S\ref{sec_trackingAlg}. There is no known theoretical reason to assume that the mass function follows a particular functional form, and as a result different authors have adopted different empirical choices (usually taking the form of some variation on a power law).  Our plan will be to try several fits and select the best model using an information criteria analysis, following the approach of \cite{li19}.

We represent the planetesimal mass function in terms of a cumulative distribution, defining $N(>M)$ to be the number of planetesimals with mass greater than $M$.  Models are usually expressed as a cumulative probability $P(>M)$, which can be trivially converted into a cumulative histogram ($N(>M)$) by multiplying by the total number of planetesimals.  When appropriate, we express some models in terms of the differential probability $p(M)$, defined such that $P(>M) = \int_{M}^{\infty} p(M^\prime) \hspace{1mm} dM^\prime$.  In general, the mass-spectra tend to have a downward curvature as $M$ increases, motivating models that are variations on a power law (tapered, truncated, piece-wise, etc.)  In this study we will consider 7 models. Below, we give the functional form and free parameters for all the models, defining the normalization coefficients in Appendix~\ref{app_prefactors}.  We refer the reader to the original works for the physical motivations and subtleties of each (also see \cite{li19} for a review). 

The Simple Power Law (SPL) has one parameter: $\alpha$. 
\begin{equation}
P(>M) = \bigg(\frac{M}{M_{p,{\rm min}}} \bigg)^{-\alpha},
\end{equation}
where $M_{p,{\rm min}}$ is the minimum mass planetesimal found in the simulation.  While this is usually not a particularly good fit to the simulated mass function, it allows for a direct comparison to several prior studies in which it was assumed \citep{abod18, simon16, schafer17, simon17_streaming}.  This fit to the cumulative distribution is the same as a simple power law fit to the differential distribution with an index of $q = \alpha+1$.\footnote{$q$ here is equivalent to $p$ in our previous works (e.g., \citealt{simon16,simon17_streaming,abod18}).}

The Simply Tapered Power Law (STPL, \citealt{abod18}) has 2 parameters: $\alpha$ and $M_{\text{exp}}$,
\begin{equation}
P(>M) = c_1 M^{-\alpha} \exp{\bigg(-\frac{M}{M_{\text{exp}}}\bigg)}.
\end{equation}

The Variably Tapered Power Law (VTPL, \citealt{schafer17}) has 3 parameters: $\alpha$, $\beta$, and $M_{\text{exp}}$,
\begin{equation}
P(>M) = c_2 M^{-\alpha} \exp{\bigg[-\bigg(\frac{M}{M_{\text{exp}}}\bigg)^{-\beta}\bigg]}.
\end{equation}

The Broken Cumulative Power Law (BCPL, \citealt{li19}) has 3 parameters: $\alpha_1$, $\alpha_2$, and $M_{\text{br}}$,
\begin{equation}
P(>M) = 
\begin{cases}
c_{31} M^{-\alpha_1} &  M \leq M_{\text{br}} \\
c_{32} M^{-\alpha_2} &  M > M_{\text{br}}.
\end{cases}
\end{equation}

The Truncated Power Law (TPL, \citealt{schafer17}) has 2 parameters: $\alpha$ and $M_{\text{tr}}$,
\begin{align}
P(>M) = 
\begin{cases}
c_4 M^{-\alpha-1} &  M \leq M_{\text{tr}} \\
0 & M > M_{\text{tr}}.
\end{cases}
\end{align}

The Broken Power Law (BPL, \citealt{li19}) has 3 parameters: $\alpha_1$, $\alpha_2$, and $M_{\text{br}}$,
\begin{align}
p(M) &= 
\begin{cases}
c_{51} M^{-\alpha_1-1} &  M \leq M_{\text{br}} \\
c_{52} M^{-\alpha_2-1} &  M > M_{\text{br}}.
\end{cases}
\end{align}

The Three Segment Power Law (TSPL, \citealt{li19}) has 5 parameters: $\alpha_1$, $\alpha_2$, $\alpha_3$, $M_{\text{br1}}$, $M_{\text{br2}}$.
\begin{align}
p(M) &= 
\begin{cases}
c_{61} M^{-\alpha_1-1} & M \leq M_{\text{br},1} \\
c_{62} M^{-\alpha_2-1} & M_{\text{br},1} < M \leq M_{\text{br},2} \\
c_{63} M^{-\alpha_3-1} & M > M_{\text{br},2}. \\
\end{cases}
\end{align}

\noindent
In all of these equations, the ``$c$" co-efficients are normalization constants set by the condition $P(>M_{p,{\rm min}}) = 1$; see Appendix B.

We fit these models to the simulation data using a maximum likelihood approach, using the fitting method, uncertainty determination, and fit evaluation methods described in \citet{li19}. The likelihood functions given in \cite{li19} allow us to determine the likelihood $\mathcal{L}$ of a model with some given set of parameters denoted as $\boldsymbol{\theta}$.  In order to find parameters of maximum likelihood, we attempt to minimize $-\text{ln}(\mathcal{L})$.  While straightforward optimization methods like gradient descent work well to find local minima, the likelihood functions for many of these models can have multiple local minima and are not necessarily smooth.  These features make the optimization problem more difficult.  An effective method in these difficult cases is MCMC, which we utilize to generate a global map of the posterior distribution of the likelihood function and give an estimate of the optimized model parameters: $\boldsymbol{\theta_{\text{MCMC}}}$.  $\boldsymbol{\theta_{\text{MCMC}}}$ is then used as an initial guess for the ``minimize" routine within Scipy's ``optimize" package, which is able to find a precise local minimum near the initial guess and finally gives us the parameters of maximum likelihood. 

We use a non-parametric bootstrapping method to determine the uncertainties in the parameters for each simulation \citep{efron94, burnham02}.  The clump masses that PLAN outputs are randomly re-sampled 1000 times with replacement such that each sample has the same number of clumps as the original sample.  Models are then fit to each of the re-samplings using the parameter optimization scheme described above, generating a 1000 point distribution for each parameter.  The $1\sigma$ confidence interval is then defined to be the region between the 16th percentile and 84th percentile of this distribution for each parameter.    

\begin{figure*}[t]
\centering
\includegraphics[width=0.8\columnwidth]{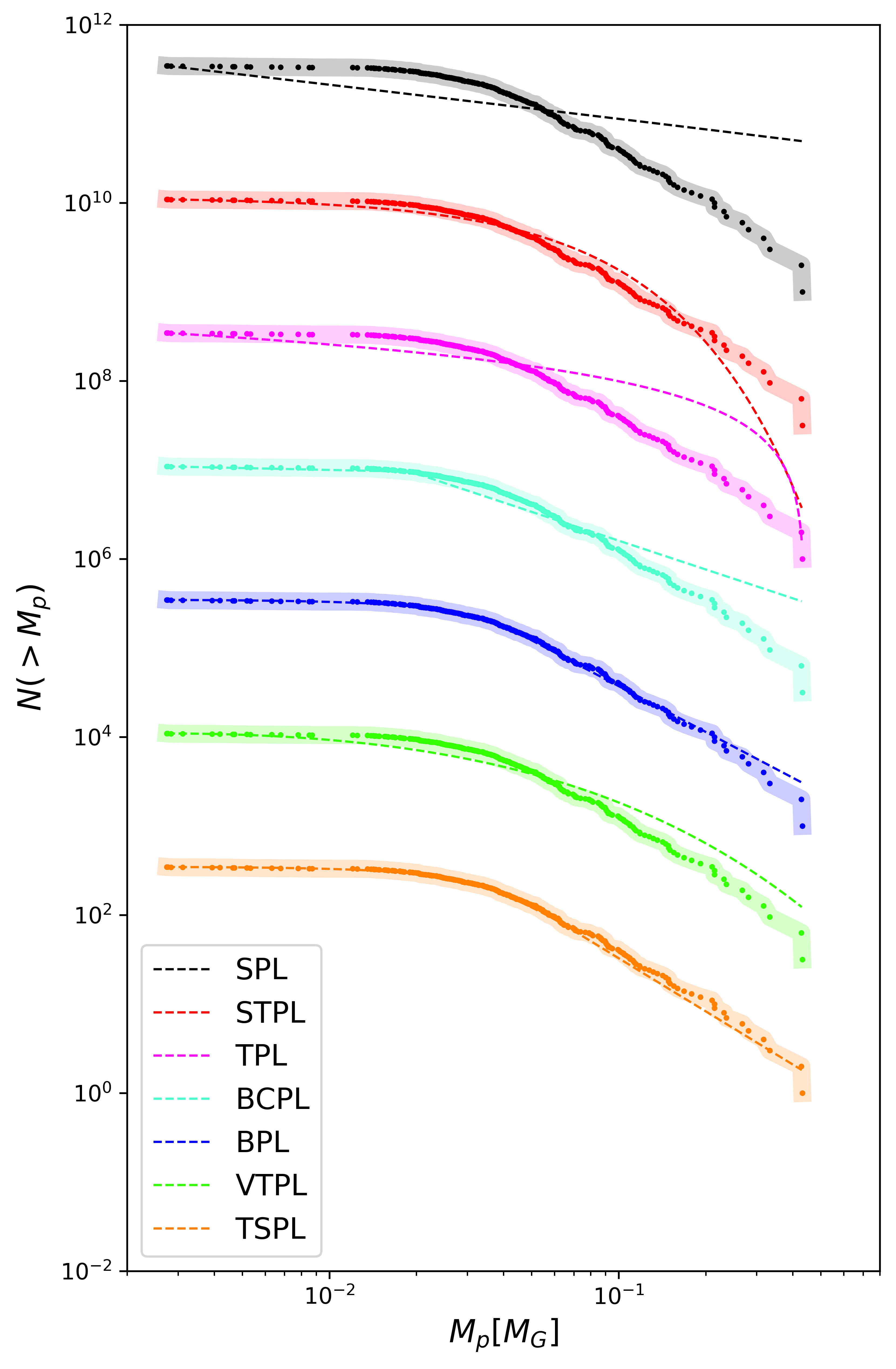}
\includegraphics[width=0.8\columnwidth]{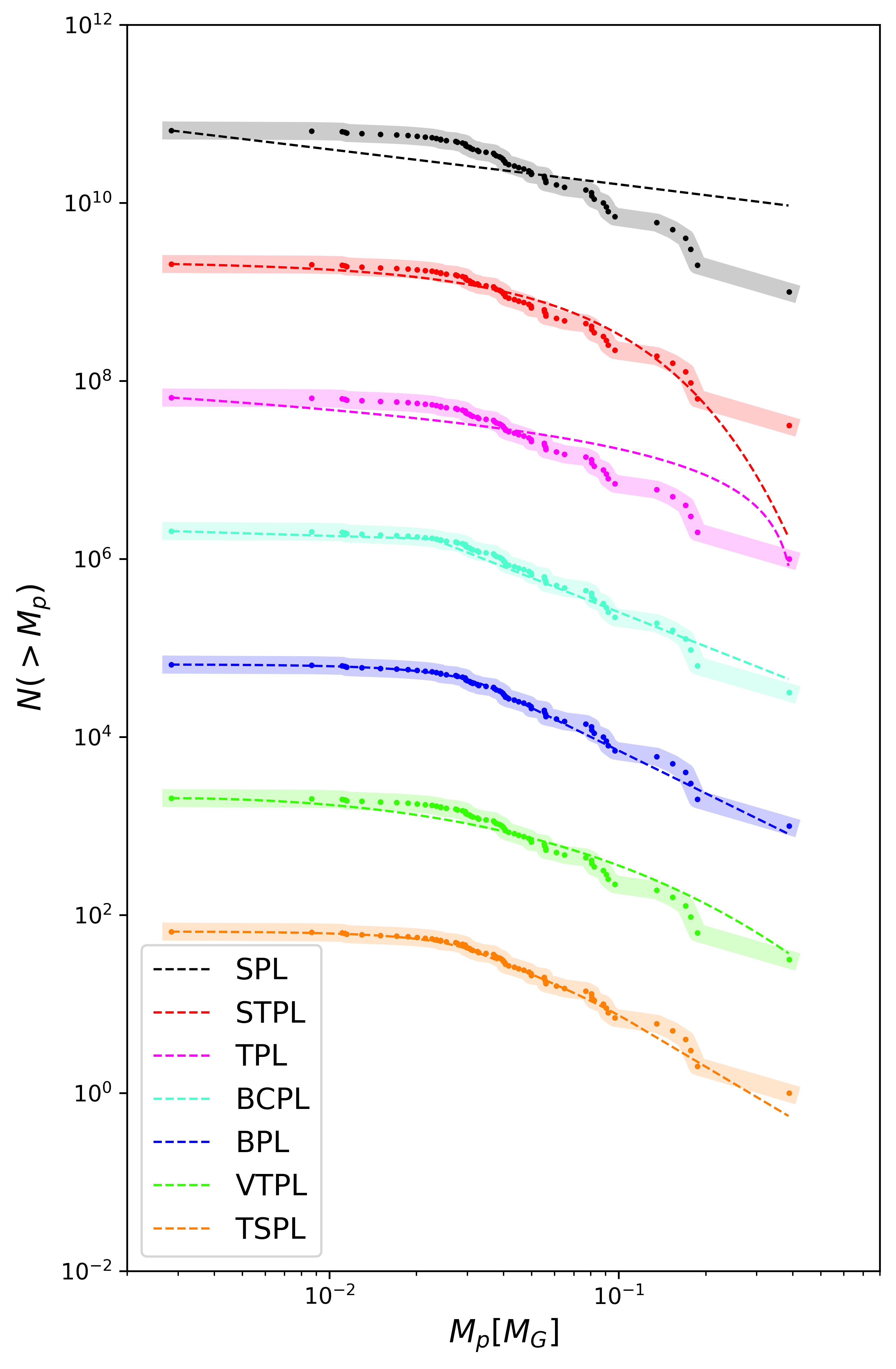}
\caption{Cumulative mass histograms and models for the control (left) and moderate turbulence (right) simulations and for 7 different models: simple power law (SPL), simply tapered power law (STPL), truncated power law (TPL), broken cumulative power law (BCPL), broken power law (BPL), variably tapered power law (VTPL), and the three segment power law (TSPL).  Fit parameters to the best models can be found in Table~\ref{tab_fits}.  The bottom histogram shows the true values, with each plot above offset by a factor of $10^{1.5}$ for convenient visualization. The shape of the planetesimal mass function is generally the same in the control (pure SI) simulation as in the moderate turbulence simulation. The broken power law model for the mass function, shown in dark blue, is statistically favored according to the Bayesian and Akaike information criteria for all of our simulations.}
\label{fig_histCumulative}
\end{figure*}

We evaluate the goodness of fit of the various models using Bayesian and Akaike information criteria \citep{kass95, akaike74}.  These criteria account for the number of fit parameters, giving an advantage to simpler models.  They are defined as follows:
\begin{align}
\text{BIC} &= K \ln{N} - 2 \ln{\mathcal{L}}, \\
\text{AIC} &= 2K - 2 \ln{\mathcal{L}},
\end{align}
where $K$ is the number of parameters in the model, $N$ is the number of data points (clump masses in this case), and $\mathcal{L}$ is the likelihood of the maximum likelihood model.  These values do not carry much meaning in isolation, as the likelihood values are affected by arbitrary constants and the sample size.  To define a metric that has physical meaning, we measure the differential between the model with the minimum BIC/AIC and all other models, defined as
\begin{align}
\Delta_{\text{BIC}} &= \text{BIC} - \text{BIC}_{\text{min}}, \\
\Delta_{\text{AIC}} &= \text{AIC} - \text{AIC}_{\text{min}}.
\end{align}
With this definition, the best model will have $\Delta_{\text{BIC}}=0$ and/or $\Delta_{\text{AIC}}=0$.  In most cases we find that both criteria identify the same best model.  The other models will have some positive value for these criteria, with higher values corresponding to worse models.  While these criteria are somewhat abstract, as a general rule we consider there to be at least some evidence for models with $\Delta_{\text{BIC}}$($\Delta_{\text{AIC}}$) between 0 and 10, and essentially no evidence for models with values greater than 10.  For a more nuanced interpretation of these criteria, we refer the reader to \cite{burnham02}. 

The cumulative spectra and fits are shown in Figure~\ref{fig_histCumulative} for initial planetesimal masses determined by a combination of PLAN and our clump tracking algorithm.  We report the best-fit parameters and information criteria for each model in Table~\ref{tab_fits}. The adoption of the clump tracking analysis has a significant impact on the results. Comparing clump tracking to an analysis of snapshots, in the case of the control simulation, we find that the average clump mass is decreased by a factor of 3 or 4, depending on which snapshot is chosen.  The characteristic masses found by the models change by similar factors as well.  This suggests that previous studies have over-predicted the true \textit{initial} masses of planetesimals formed by the SI/SG mechanism. In addition, because our clump-tracking algorithm removes splitter clumps (see Case B in \S\ref{sec_trackingAlg}), we find a shallower power-law slope in the cumulative distribution at the low-mass end.

In this study, information criteria favor the broken power law model in every case. The values of $\alpha_1$ measured from this fit imply that a turnover is observed in all runs. The three segment power law is often a close second if the AIC -- which is less punishing towards models with more parameters -- is considered. Both of these favored models are broken power laws in the differential distribution, translating to a smooth transition between power law indices in the cumulative distribution.   

In comparison to \cite{abod18}, who fit the STPL to their distribution, we generally find that our characteristic masses are lower by a factor of a few and that our power law index is significantly lower at low-masses. These effects are primarily due to the adoption of the clump tracking analysis. If instead we analyze snapshots from both our control and driven-turbulence simulations, we find that our model parameters are generally consistent with prior studies.

\begin{deluxetable*}{lllllllll}
\rotate
\tablecaption{Best Fit Parameters \label{tab_fits}}
\tablewidth{1.8\columnwidth}
\tablehead{ \colhead{Label} & \colhead{Model} & \colhead{$\Delta_{\text{BIC}}$} & \colhead{$\Delta_{\text{AIC}}$} & \colhead{Param 1} & \colhead{Param 2} & \colhead{Param 3} & \colhead{Param 4} & \colhead{Param 5} } 
\startdata
Control      &    SPL      &    545       &    564       &    $\alpha    = 0.3743 ^{+0.0063 }_{-0.0063 }$ \\
             &    STPL     &    59.6      &    69.3      &    $\alpha    = 0.0025 ^{+0.0005 }_{-0.0003 }$    &    $M_{exp}   = 0.0527^{+0.0031 }_{-0.0032 }$ \\
             &    TPL      &    305       &    314       &    $\alpha    = 0.0101 ^{+0.0023 }_{-0.0022 }$    &    $M_{tr}    = 0.433^{+0.003 }_{-0.051 }$ \\
             &    BCPL     &    83.4      &    83.4      &    $\alpha_1  = 0.072 ^{+0.021  }_{-0.013 }$      &    $\alpha_2  = 1.107^{+0.122 }_{-0.073 }$        &    $M_{br}    = 0.0195 ^{+0.0038 }_{-0.0014 }$ \\
             &    BPL      &    0.0       &    0.0       &    $\alpha_1  = -1.57^{+0.20 }_{-0.16 }$          &    $\alpha_2  = 1.71^{+0.18 }_{-0.13 }$           &    $M_{br}    = 0.0412 ^{+0.0043 }_{-0.0031 }$ \\
             &    VTPL     &    128       &    128       &    $\alpha    = -0.4169^{+0.0085 }_{-0.0088 }$    &    $\beta     = 0.3871 ^{+0.0067 }_{-0.0056 }$    &    $M_{exp}   = 0.0034 ^{+0.0001 }_{-0.0    }$ \\
             &    TSPL     &    20.5      &    1.1       &    $\alpha_1  = -1.65^{+0.17  }_{-0.38 }$         &    $\alpha_2  = 0.88 ^{+0.48   }_{-1.19  }$       &    $\alpha_3  = 2.14 ^{+0.93 }_{-0.25 }$    &   $M_{br1}   = 0.0358 ^{+0.0041  }_{-0.010 }$    &    $M_{br2}   = 0.07 ^{+0.11 }_{-0.02 }$ \\
\hline
Weak         &    SPL      &    140       &    155       &    $\alpha    = 0.374  ^{+0.071 }_{-0.019 }$ \\
             &    STPL     &    47.3      &    54.5      &    $\alpha    = 0.018 ^{+0.053 }_{-0.010 }$       &    $M_{exp}   = 0.066 ^{+0.041 }_{-0.022 }$ \\
             &    TPL      &    115       &    122       &    $\alpha    = 0.142  ^{+0.074 }_{-0.128 }$      &    $M_{tr}    = 1.66 ^{+0.65 }_{-1.43 }$ \\
             &    BCPL     &    7.8       &    7.8       &    $\alpha_1  = 0.093 ^{+0.028 }_{-0.020 }$       &    $\alpha_2  = 1.23 ^{+0.15  }_{-0.15 }$         &    $M_{br}    = 0.0262 ^{+0.0019 }_{-0.0105 }$ \\
             &    BPL      &    0.0       &    0.0       &    $\alpha_1  = -1.54^{+0.38 }_{-0.45 }$          &    $\alpha_2  = 1.48 ^{+0.37 }_{-0.21 }$          &    $M_{br}    = 0.038 ^{+0.012 }_{-0.014 }$ \\
             &    VTPL     &    39.4      &    39.4      &    $\alpha    = -0.44^{+0.027 }_{-0.11 }$         &    $\beta     = 0.341 ^{+0.028 }_{-0.033 }$       &    $M_{exp}   = 0.0063 ^{+0.0084 }_{-0.0015 }$ \\
             &    TSPL     &    16.3      &    1.8       &    $\alpha_1  = -1.58^{+0.30 }_{-0.53  }$         &    $\alpha_2  = 0.80 ^{+0.36 }_{-0.79  }$         &    $\alpha_3  = 1.67 ^{+1.47 }_{-0.30 }$    &    $M_{br1}   = 0.0321 ^{+0.0058 }_{-0.0139 }$    &    $M_{br2}   = 0.073  ^{+0.058 }_{-0.026 }$ \\
\hline
Moderate     &    SPL      &    89.1      &    102       &    $\alpha    = 0.399 ^{+0.289 }_{-0.021 }$ \\
             &    STPL     &    9.5       &    15.8      &    $\alpha    = 0.018 ^{+0.058 }_{-0.006 }$       &    $M_{exp}   = 0.050 ^{+0.011 }_{-0.033 }$ \\
             &    TPL      &    51.1      &    57.5      &    $\alpha    = 0.05 ^{+0.29 }_{-0.02 }$          &    $M_{tr}    = 0.20 ^{+0.22 }_{-0.09 }$ \\
             &    BCPL     &    9.8       &    9.8       &    $\alpha_1  = 0.12 ^{+0.14 }_{-0.04 }$          &    $\alpha_2  = 1.42 ^{+0.23 }_{-0.18 }$          &    $M_{br}    = 0.023 ^{+0.005 }_{-0.015 }$ \\
             &    BPL      &    0.0       &    0.0       &    $\alpha_1  = -1.57^{+0.38 }_{-0.51 }$          &    $\alpha_2  = 1.64 ^{+0.32 }_{-0.22 }$          &    $M_{br}    = 0.034 ^{+0.008 }_{-0.022 }$ \\
             &    VTPL     &    26.5      &    26.5      &    $\alpha    = -0.464^{+0.028 }_{-0.306 }$       &    $\beta     = 0.362 ^{+0.097 }_{-0.023 }$       &    $M_{exp}   = 0.0071 ^{+0.0056 }_{-0.001 }$ \\
             &    TSPL     &    16.5      &    3.8       &    $\alpha_1  = -1.71^{+0.49 }_{-0.66 }$          &    $\alpha_2  = 0.83  ^{+0.37 }_{-1.52 }$         &    $\alpha_3  = 2.04 ^{+1.39 }_{-0.40 }$    &    $M_{br1}   = 0.0311 ^{+0.0082 }_{-0.0182 }$    &    $M_{br2}   = 0.064  ^{+0.069 }_{-0.046 }$ \\
\enddata
\tablecomments{Summary of fits to the cumulative mass function of planetesimals based on the initial masses of each clump from our clump-tracking analysis.}
\end{deluxetable*}

\subsection{The effect of turbulence}
One of the motivating questions for this study was whether or not turbulence changes the mass-function of planetesimals.  If we compare like-models across the 3 simulations, we see that the distributions are generally consistent with each other. The strength of this conclusion is limited, in part, by the relatively large errors on the parameters resulting from the low sample size, in particular in the case of moderate turbulence. However, our results in the presence of turbulence provide limited additional evidence in support of the idea that the shape of the initial mass function resulting from gravitational collapse is ``universal" \citep{simon17_streaming}. (Note however that \citealt{li19} find evidence against this hypothesis.) While the number of clumps that form may change, the shape of the mass-function does not. 

\begin{figure}[h]
\centering
\includegraphics[width=0.98\columnwidth]{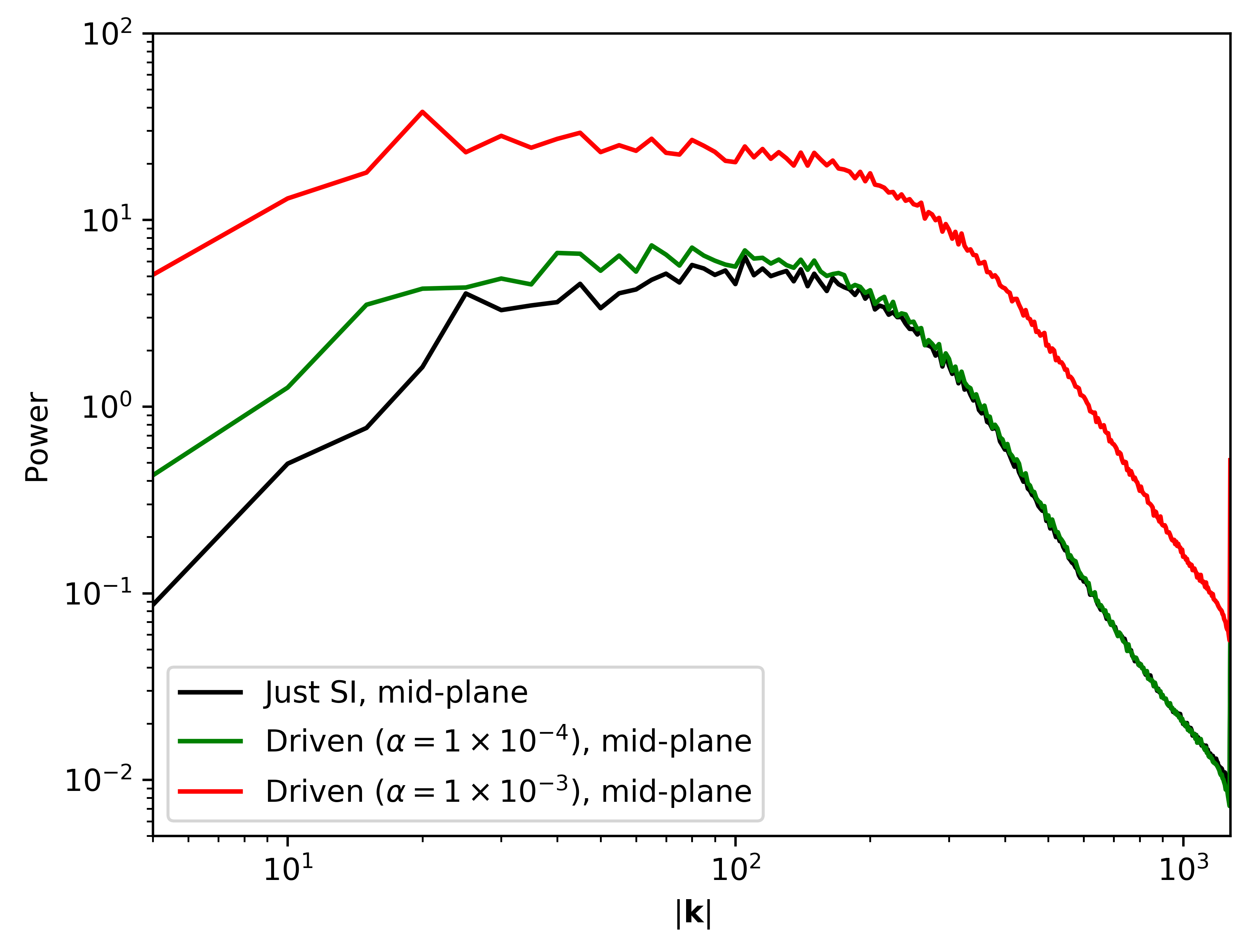}
\caption{The power spectrum of the fluid velocity perturbations within a small slice at the mid-plane for 3 runs: control (black, $\alpha_{\rm drive}=0$), weak driving (green, $\alpha_{\rm drive}=10^{-4}$), and strong driving (red, $\alpha_{\rm drive}=10^{-3}$).  $|\mathbf{k}|$ is given in code units ($1/H$).  These spectra are compensated such that power law with an index of $-5/3$ would appear to be flat.  They are normalized by a common factor such that the mean of the control curve is unity.  Using this convention the absolute values of the power bear no meaning, but the relative power between the spectra is maintained.  The weak driving and control runs have very similar spectra at small scales but differ on larger scales, which is consistent with small scale motions being dominated by the presence of a thin particle layer.}
\label{fig_ps_compare}
\end{figure}

In considering the influence of intrinsic turbulence on planetesimal formation, it is important to bear in mind that the streaming instability itself drives some level of fluid turbulence. In an attempt to understand the physical mechanism for driven turbulence's effect on the rate of planetesimal formation, we compare the fluid velocity power spectrum of SI driven turbulence in our control experiment to the power spectrum of our driven turbulence.  This comparison is shown in Fig.~\ref{fig_ps_compare}, with the power spectra being calculated according to the scheme given in equations \ref{eq_ps1} through \ref{eq_ps3} and using the compensation method described in Appendix~\ref{app_stirTurb}.

On small size-scales ($|\mathbf{k}| \gtrsim 100$), the weakly driven turbulence has a nearly identical spectrum and amplitude as the turbulence from the streaming instability itself.   On the largest scales, the driven turbulence shows more power than SI alone.  This can be understood intuitively, as our driving scheme distributes power across all scales by design, whereas the SI would generally be expected to stir motions on smaller scales.  The run with stronger driven turbulence unsurprisingly has more power across all scales.

In general, one would expect that the collapse of particles into gravitationally bound clumps is primarily influenced by fluid motions on the scale of or smaller than the Toomre unstable wavelength (equation \ref{eq_toomre}; see e.g. \citealt{abod18}).  For our choice of parameters, this length scale corresponds to $0.016H$ -- or equivalently about $k=62$.  On this scale and smaller (higher $k$), the pure SI and weakly driven simulations have very similar power spectra.  Given this similarity on the relevant size-scales, it is not surprising that the two simulations form planetesimals with consistent initial mass distributions.  However, the discrepancy in the formation rate between these two simulations (Table \ref{tab_highLevel}) is less obvious to explain.  We speculate that this is due to a difference in the particle density at the mid-plane.  Most of the energy in Kolmogorov turbulence is on large scales.  As a result, $v_z$ modes at large scales will be the primary source of vertical diffusion of particles, and the spectra so differ on these scales.  Consequently, the particle layers will have a different width and density. We find that the mid-plane particle density in the control simulation is about twice that of the weak turbulence simulation.  This difference is equivalent to changing the {\it local} solid-to-gas ratio, $\epsilon$, which is known to strongly effect the SI's behavior \citep{youdin07b}.

Our results support the idea that turbulence influences planetesimal formation by the SI/SG mechanism primarily through its effect on the thickness of the particle layer, which changes the mid-plane value of $\epsilon$. The mid-plane solid-to-gas ratio can be approximated as
\begin{equation}
\label{eps_eqn}
    \epsilon = Z \bigg( \frac{h_{\text{gas}}}{h_\text{par}} \bigg) \approx Z \bigg( \frac{\tau}{\alpha_{\text{turb}}} \bigg)^{1/2},
\end{equation}
where the approximate equality holds when particles are well-coupled to the gas. For our parameter choices ($Z=0.02$, $\tau=0.3$), we find that the dividing line between turbulence that admits and precludes planetesimal formation occurs at about $\alpha_{\text{turb}} = 10^{-3.25}$ (halfway between the two cases tested). This corresponds to an approximate critical $\epsilon$ value $\epsilon_{\text{crit}} = 0.46$.

\begin{figure}[h]
\centering
\includegraphics[width=0.98\columnwidth]{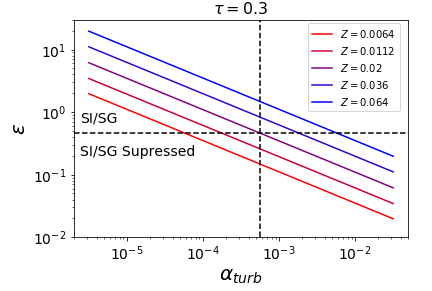}
\caption{The mid-plane solid-to-gas ratio $\epsilon$ shown as a function of $\alpha_\text{turb}$ for various values of $Z$, according to our very simple model that is applicable when particles are well bound to the gas.  The vertical and horizontal dashed lines show our empirically found critical $\alpha$ and $\epsilon$ values respectively.  We hypothesize that combinations of $\alpha_{\text{turb}}$ and $Z$ that lie below the horizontal dashed line will produce insufficient particle densities at the mid-plane to form planetesimals via the SI/SG mechanism. We emphasize that this figure is generated based on a single, empirically determined value from our study in combination with equation~\ref{eps_eqn} and should be considered a toy model that extends our results under some naive assumptions.}
\label{fig_alpha_epsilon}
\end{figure}

The $\epsilon$ threshold that we have derived is broadly consistent with the analytic result that SI growth rates are much faster for $\epsilon \geq 1$ (see \cite{youdin05} Figure 1). Based on this, we hypothesize that our results could be extended to different values of the $Z$ parameter. As shown in Figure~\ref{fig_alpha_epsilon}, such an extrapolation would imply that in truly strong disk turbulence ($\alpha_{\rm turb} \sim 10^{-2}$), planetesimal formation would only be possible given $Z \gtrsim 0.1$. Although one might be tempted to extrapolate further to consider variations in $\tau$, the differing aerodynamic response as $\tau$ is varied makes such an extrapolation less secure.

\section{Summary \& Discussion}
\label{sec_discuss}
In this paper we have presented simulations of planetesimal formation in the presence of intrinsic disk turbulence, using a small local domain to attain high spatial resolution, and including self-gravity to follow gravitational collapse. Our primary results are as follows.

\begin{enumerate}
\item For our fiducial parameters ($Z=0.02$, $\tau=0.3$), turbulence at the level of $\alpha_{\text{drive}}=10^{-3}$ does not permit the formation of gravitationally bound particle clumps via the SI/SG mechanism.  We interpret this result as being due to the increased width of the particle layer stirred by the turbulence, which results in lower particle densities at the mid-plane.  The lower densities prevent the streaming instability from being able to concentrate particles to a sufficient density for clumps to be bound by their own self gravity.  For our parameter choices, the threshold of turbulence above which planetesimals do not form is somewhere between $\alpha_{\text{drive}}=10^{-3.5}$ and $\alpha_{\text{drive}}=10^{-3}$, or equivalently $\delta v / c_s=0.018$ and $0.032$. The inferred critical solid-to-gas ratio at the mid-plane is $\epsilon_{\rm crit} \simeq 0.5$.
\item Intermediate turbulence ($\alpha_{\text{drive}}=10^{-4}$ to $10^{-3.5}$) permits the formation of planetesimals, but they form more slowly than in a control run with no imposed turbulence.
\item Under the hypothesis that the threshold for planetesimal formation corresponds to a mid-plane solid-to-gas ratio $\epsilon > \epsilon_{\rm crit}$, we can extrapolate the simulation results to different $Z$ (Figure~\ref{fig_alpha_epsilon}). At $\alpha_{\rm turb} = 10^{-2}$, for example, planetesimal formation would not occur unless $Z \gtrsim 0.1$.
\item  We have developed a new clump-tracking method that allows us to measure the true initial masses of planetesimals.  This method reduces the potentially confounding post-formation effects (e.g., mergers, accretion, tidal stripping) present when one measures masses of all planetesimals in a single snapshot.  This method yields lower masses than found in previous work (which used snapshots) by a factor of $\sim3$, and we find that the slope of the mass function at low masses is shallower than in those prior analyses. 

\item The initial mass function of planetesimals is well described by a broken power law model, and the best fit parameters of that model are robust to the presence and strength of imposed turbulence.

\end{enumerate}

The results we have presented here are subject to several uncertainties.  First, by ``manually" imposing turbulence onto our domain and not including self-consistently driven turbulence, we may be neglecting important differences in the structure of the turbulent fluctuations.  For example, our turbulent stirring mechanism treats all three dimensions equally, whereas the MRI produces turbulent fluctuations that are anisotropic \cite[e.g.,][]{hawley95,guan09}.
Second, in this study we have only varied one parameter describing the strength of turbulence. It remains to be seen how the $\alpha_{\text{drive}}$ threshold for planetesimal formation changes with varying $\tau$, $Z$, $\tilde{G}$, and $\Pi$. 

Furthermore, our results are in apparent conflict with studies by \cite{schafer20}, \cite{johansen07}, and \cite{yang18},
which showed that planetesimal formation is possible in turbulent environments. \cite{schafer20} demonstrated this in the presence of turbulence due to the vertical shear instability (VSI), whereas the latter two papers made a similar determination in the presence of MRI-generated turbulence.  In particular, \cite{johansen07} obtained bound planetesimals despite MRI driven turbulence (in the ideal MHD limit) at a level $\alpha_{\text{turb}} \sim 10^{-3}$. \cite{yang18} showed that modest to strong particle clumping occurs in the presence of ideal MHD and in Ohmic dead zones, though without self-gravity it is unclear which of the \cite{yang18} runs would produce planetesimals, and at what rate.

The differences between our results and these works may be attributable to different regions of parameter space, as we discussed above.  However, we should also emphasize that the driven turbulence in our simulations is not interchangeable with VSI or MRI turbulence.  The planetesimal formation that proceeds in the VSI simulations of \cite{schafer20} result from large-scale overdensities that seed the SI.  In the case of planetesimal formation under MRI turbulence, zonal flows and other large-scale effects (e.g., anisotropic diffusion of particles) are able to seed the SI on large scales \citep{johansen07,yang18}. We hypothesize that the primary differentiating effect between these works and the results presented here are the large-scale overdensities produced by the MRI and VSI.  Absent these overdensities, we find that turbulence produces a hindering effect on planetesimal formation.  Given the wide array of turbulence-inducing mechanisms in disks, we feel that isolating the effects of simple isotropic turbulence on the SI and planetesimal formation is a worthwhile exercise and a useful result.  Furthermore, in Fourier space, the rate at which overdensities are diffused via turbulence scales with the wavenumber as $k^2$.  Thus, resolving smaller scales than the aforementioned work, where turbulent diffusion may dominate over concentrating effects, is an important endeavor to pursue. Ideally, one would like to study planetesimal formation including both the high resolution used here, and the large domain of \cite{yang18}. This aspiration is not feasible with fixed-grid simulations, but may be possible using mesh refinement methods. Ultimately, while our approach by no means completely solves the question of how planetesimal formation proceeds in turbulent environments, our results add another piece to the puzzle, which we hope will motivate and inform future work along these lines.

These considerations aside, the precise implications of our results for planet formation depend upon the radial variation of $\alpha_{\rm turb}$ and $Z$ in protoplanetary disks. These functions are poorly constrained observationally, and many physical effects (including radial drift, the possibility of zonal flows, and different sources of turbulence) hamper their calculation from first principles. At large orbital radii, however, there is both theoretical and observational evidence to suggest that turbulence is weak. At these distances, strong ambipolar diffusion \citep{kunz04,desch04} results in weak MHD turbulence near the mid-plane, with $\alpha_{\text{turb}} \sim 10^{-4}$ \citep{bai15,simon13a,simon13b}. These values are consistent with observational constraints at distances larger than $\sim$ 30~AU away from the central star \citep{pinte16,flaherty17}, and would be low enough to allow for planetesimal formation according to the criterion suggested by our simulations. The same is probably true on intermediate scales $\sim$1--30AU, though the complexity of the MHD physics at radii where both Ohmic diffusion and the Hall effect are important adds some uncertainty. 

Our criterion predicts that planetesimal formation would not be possible in the extreme inner regions of protoplanetary disks, interior to about 0.1~AU, except perhaps at very high $Z$. At radii where the mid-plane temperature exceeds $T \simeq 10^3 \ {\rm K}$, ionization of the alkali metals is sufficient to allow the MRI to operate under effectively ideal MHD conditions \citep{gammie96}. The predicted values of $\alpha$ on these small scales are estimated to be $\alpha_{\text{turb}} \sim 10^{-2}$, or possibly greater if a strong net-vertical-field exists \citep{simon13b, salvesen16}. Our results would then suggest that planetesimals will not form in this region, preventing in situ formation \citep{batygin16} of some super-Earths at an early phase. At least some degree of migration would then be favored for the formation of close-in exoplanets \citep{masset03, beauge12}.    

In summary, our results suggest that isotropic turbulence would allow planetesimal formation across most of the disk in regions of reduced ionization and turbulence, but is precluded where MRI turbulence is sufficiently vigorous, such as close to the central star.  Our findings also imply that the ability of turbulence-inducing mechanisms like the MRI and VSI to generate large-scale overdensities of particles may be an essential property for allowing planetesimal formation above the turbulent disruption threshold we have found for isotropic turbulence.  While these results need to be verified with a broader set of simulations, the tools that we have developed here and our preliminary results are encouraging for future studies of planetesimal formation in the complex environments present within protoplanetary disks.

\acknowledgements
JBS acknowledges support from NASA under {\em Emerging Worlds} through grant 80NSSC18K0597. ANY acknowledges support from NASA Astrophysics Theory Grant NNX17AK59G and from NSF grant AST-1616929.  RL acknowledges support from NASA headquarters under the NASA Earth and Space Science Fellowship Program grant NNX16AP53H. PJA acknowledges support from NASA under grants NNX16AB42G and 80NSSC19K0639.

\appendix

\section{Driven Turbulence Methods and Testing}
\label{app_stirTurb}
\setcounter{figure}{0} \renewcommand{\thefigure}{A.\arabic{figure}}

In this Appendix, we describe the methods used to drive turbulence within the simulation domain. We perform several tests to confirm that the turbulence is homogeneous, isotropic, and has the specified power spectrum.  

The amplitudes of the velocity fluctuations in $k$-space are given by
\begin{equation}
A(k_x,k_y,k_z) = (\text{rand}) (|k|)^{-b}, 
\end{equation}
where $A$ is the amplitude of the mode and ``$\text{rand}$" is a random number drawn from a normal distribution.  The power law index $b$ can be related to $p$ via dimensional arguments.  Starting with
\begin{equation}
E_1(k) \propto k^{-p},
\end{equation}
and multiplying both sides by $dk$ yields an expression with units of energy:
\begin{equation}
E_1(k)dk \propto k^{-p} dk.
\end{equation}
Since we are assigning velocity amplitudes in 3D $k$-space, we need to convert this to be a volume density ($E_3$) instead of a linear density,
\begin{equation}
E_1(k)dk \propto E_3(k_x, k_y, k_z) dk_x dk_y dk_z \propto E_3(k)(4\pi k^2)dk.
\end{equation}
By inspection, the power law index for $E_1$ must be 2 greater than the power law index for $E_3$,
\begin{equation}
E_3(k) \propto k^{-p-2}.
\end{equation}
The velocities are related to this energy by a square root:
\begin{equation}
\sqrt{E_3(k)} \propto A(k_x,k_y,k_z),
\end{equation}
where $A$ is the amplitude of a velocity mode in 3D $k$-space, which is by definition a volume-density in $k$-space.  Finally, combining the two previous proportionalities yields
\begin{equation}
A \propto k^{-b}, \hspace{3mm} b=\frac{p}{2}+1.
\end{equation}
So if we want $E_1(k) \propto k^{-5/3}$, then we need to set the velocity modes with amplitude $A \propto k^{-11/6}$.  

We assign these amplitudes on a 3D lattice for all ($k_x, k_y, k_z$) such that $|k|$ is between $k_{\text{low}}$ and $k_{\text{high}}$.  We define $k_{\text{low}}$ to be the wavenumber such that exactly one mode fits in the box and $k_{\text{high}}$ to correspond to the Nyquist wavenumber for the given resolution: $k_{\text{high}} = (N_x/2)k_{\text{low}}$.  Finally, the non-solenoidal (``divergence-full") component of the vector field is projected off, ensuring that the perturbations we are applying are incompressible.  

After these amplitudes are defined in $k$-space, the inverse Fourier transform is taken to calculate velocity perturbations with the desired spectrum.  The total energy of these perturbations is then scaled such that the energy injection rate throughout the simulation will be a constant: $({dE}/{dt})_{\text{drive}}$.  

This energy injection rate is our direct control parameter, but we really want to be able to specify the amount of turbulence in terms of the $\alpha_{\text{drive}}$ parameter, defined in equation~(\ref{eq_alpha}).  Hence, we need to measure the relationship between the energy injection rate and the observed $\delta v$ of the box.  In principle, the energy injection rate would simply balance the energy lost due to numerical dissipation.  In this case the energy injection rate--$\delta v$ scaling would match the naive guess based on an energy argument: $({dE}/{dt})_{\text{drive}} \propto (\delta v)^2$.  However, with our numerical setup there are outflows through the top and bottom boundaries of the box \citep{bai13a, ogilvie12, li18}, which add another sink for kinetic energy. The strength of those outflows of course increases with $\delta v$. As a result there is a slight non-linearity in the relation between the energy injection rate and the square of the saturated velocity fluctuations.

\begin{figure}[h]
\centering
\includegraphics[width=0.5\columnwidth]{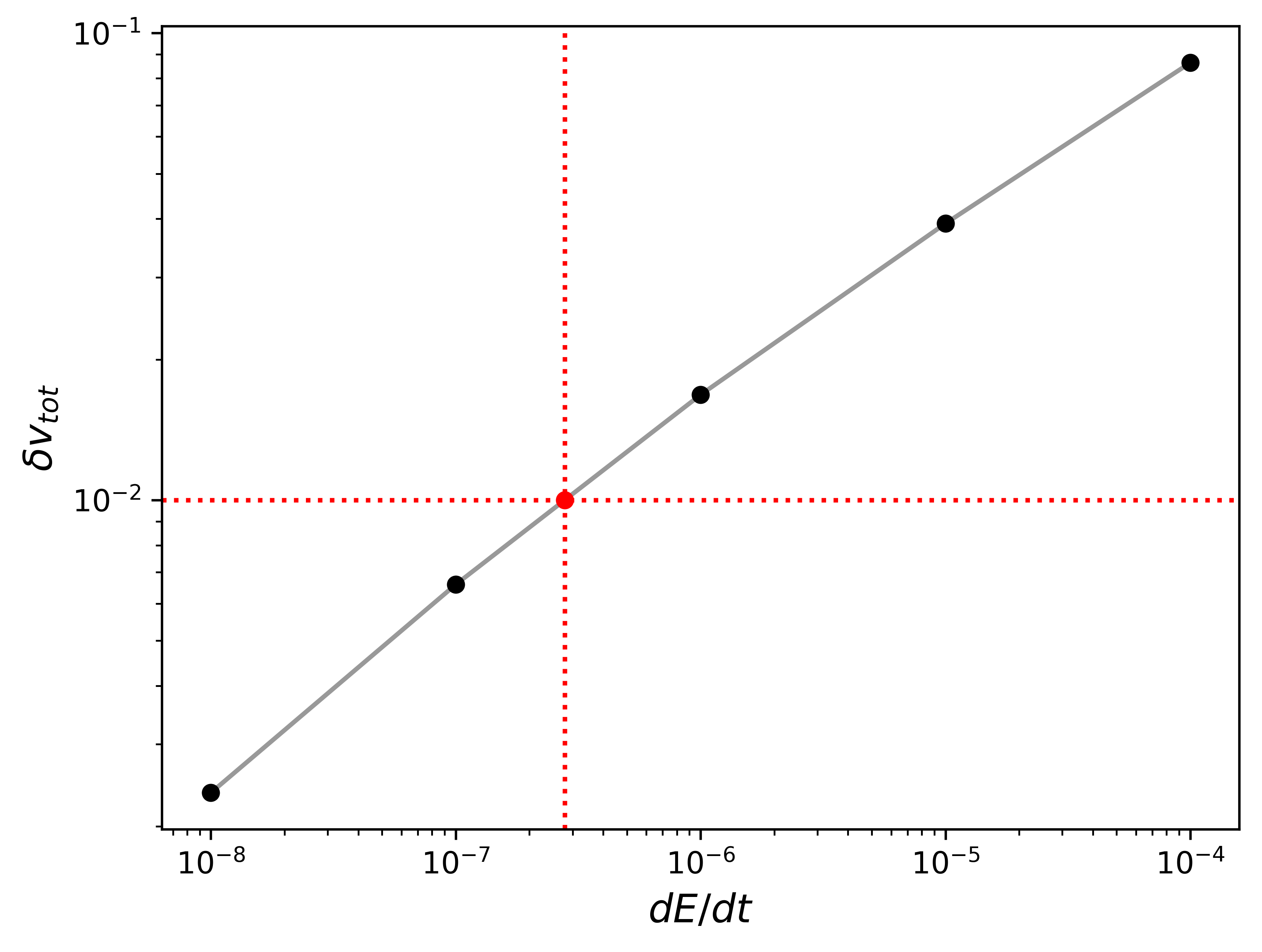}
\caption{The relationship between the driving amplitude $dE/dt$ and the resulting average velocity perturbation $\delta v$.  Each black point is generated by averaging over a low-resolution simulation.  The grey lines show a linear interpolation between the black points in log-space.  The red lines demonstrate how the appropriate driving rate is found to produce a simulation with $\delta v = 10^{-2}$, ie. $\alpha_{\text{drive}} = 10^{-4}$.}
\label{fig_5_dedv}
\end{figure}

To understand the energy injection rate--$\delta v$ scaling, we perform a series of tests at low resolution in which energy injection rates are varied by factors of 10 over 5 orders of magnitude. Once each simulation reaches a saturated state, the average $\delta v$ is calculated.  The resulting relation is shown in Figure~\ref{fig_5_dedv}. Using this data, we create a relation between the energy injection rate and the resulting velocity perturbation by linearly interpolating between these 5 data points in log-space.  This interpolation tells us the appropriate $({dE}/{dt})_{\text{drive}}$ parameter in order to achieve the desired $\delta v$ (i.e. $\alpha_{\text{drive}}$) in the saturated state of our high resolution simulations. 

Before running simulations with particles, we characterized the properties of driven turbulence in our simulated boxes to ensure that our driving algorithm met several requirements:
\begin{itemize}
\item The turbulence is isotropic. The space-time-averaged perturbations of the 3 components of the velocity are the same.
\item The turbulence is homogeneous. The time-averaged velocity perturbations are the same throughout the box.
\item The turbulence has the desired spectrum.  The velocity perturbations calculated using the methods described in \S\ref{sec_stirTurb} must translate to the intended turbulent power spectrum in the shearing box.
\end{itemize}

We performed these tests in the {\em absence} of particles for several reasons.  First, it is not obvious that the arguments leading to the Kolmogorov power spectrum hold with particles feeding back on the gas. In addition, particles tend to settle to the mid-plane, making the system inherently inhomogeneous. Finally, the streaming instability will itself drive turbulence to some degree, and it is not necessarily true that this turbulence will have a Kolmogorov spectrum.

\begin{figure}[h]
\centering
\includegraphics[width=0.5\columnwidth]{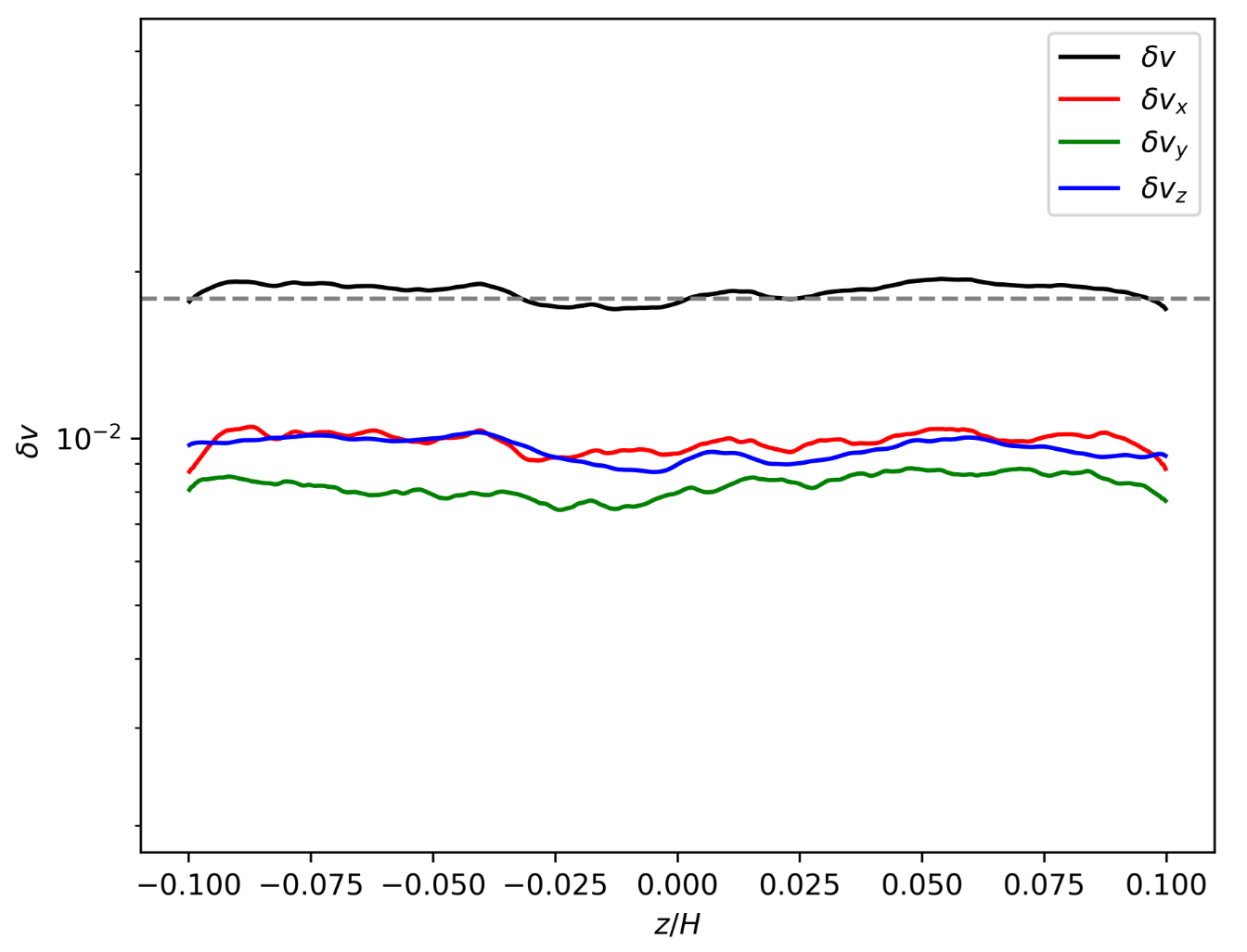}
\caption{A demonstration that our stirring algorithm produces a box of homogeneous, isotropic turbulence.  Data is from a high resolution simulation ($512^3$) with no particles.  Time averages are taken from $5 \Omega^{-1}$ to $10 \Omega^{-1}$.  The dashed horizontal line shows the expected $\delta v$ given the strength of the driving. }
\label{fig_5_stirHomoIso}
\end{figure}

Figure \ref{fig_5_stirHomoIso} shows a vertical profile of each component of the velocity perturbation, averaged in space over the horizontal directions, and in time from $t=5 \Omega^{-1}$ to $10\Omega^{-1}$. Although this is a fairly short time-average, we find that turbulence saturates quite quickly given our driving scheme. The profile for each component is relatively flat with respect to the vertical direction (shown in the figure), as well as the horizontal directions, confirming that our driving scheme produces homogeneous turbulence.  The three components of the velocity are also roughly equal throughout the box, confirming isotropy, with the possible exception of $\delta v_y$. It is plausible that this anisotropy in the $y$-direction (at about $\sim 10\%$ level) is due to the shear velocity in the box, which is also in $y$-direction and which introduces an inherently anisotropic effect.  

To calculate the perturbed kinetic energy power spectrum of the fluid we follow the methods described in \cite{simon09}.  The Fourier transform of a quantity $f(\mathbf{x})$ is represented by $\widetilde{f}(\mathbf{k})$ and is defined as usual,
\begin{equation}
\label{eq_ps1}
\widetilde{f}(\mathbf{k}) = \int \int \int f(\mathbf{x}) e^{-i \mathbf{k} \cdot \mathbf{x}} d^3 \mathbf{x}.
\end{equation}
The kinetic energy density power spectrum is then calculated as,
\begin{equation}
\label{eq_ps2}
\text{PS}\bigg(\frac{1}{2}\rho |\mathbf{\delta v}|^2 \bigg) (\mathbf{k}) = \frac{1}{2} \sum_{i=x,y,z}^{} |\widetilde{\sqrt{\rho} \delta v_i (\mathbf{k})}|^2. 
\end{equation}
In order to arrive at a spectrum that represents a linear density in $k$-space, we integrate the power over spherical shells of constant $|\mathbf{k}|$,
\begin{equation}
\label{eq_ps3}
\text{PS}\bigg(\frac{1}{2}\rho |\mathbf{\delta v}|^2 \bigg) (k) = \int \text{PS}\bigg(\frac{1}{2}\rho |\mathbf{\delta v}|^2 \bigg) (\mathbf{k}) \hspace{2mm} 4 \pi k^2 dk.
\end{equation}
Here, the convention is that $k = |\mathbf{k}|$.  We will plot a ``compensated" version of this power spectrum \citep{lemaster09}, in which the power is multiplied by $k^p$, where $p$ is the expected power law index of the spectrum ($p=5/3$ in our case).  When plotted in this way the spectra appear flat if following the power law and any deviations from this power law become more obvious.  After compensation, we normalize the spectra such that the average is one.  Note that this normalization is different than the convention used in Figure~\ref{fig_ps_compare}.

\begin{figure}[h]
\centering
\includegraphics[width=0.5\columnwidth]{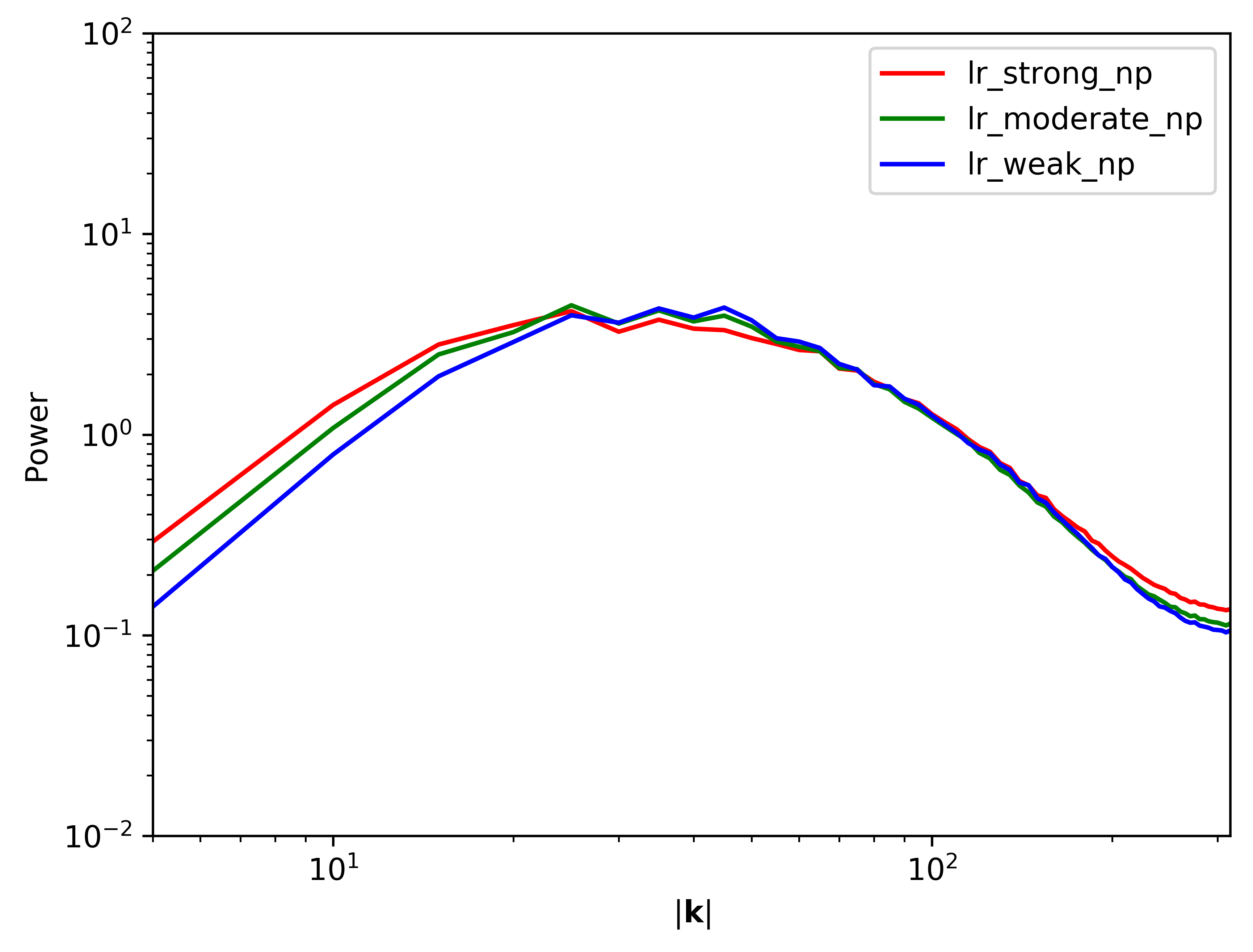}
\caption{A demonstration that our stirring algorithm produces roughly the same spectrum for three different levels of driving, and that the spectrum roughly follows a $p=-5/3$ power law over the inertial range. $|\mathbf{k}|$ is given in code units ($1/H$).  The power plotted here is normalized such that the mean is one for each of the three spectra shown.  The power is also compensated with a $k \propto -5/3$ dependence, such that a $-5/3$ power law would appear perfectly horizontal on this plot. Time averages are taken from $10 \Omega^{-1}$ to $50 \Omega^{-1}$.}
\label{fig_5_stirSpec}
\end{figure}

The power spectra for simulations without particles can be seen in Figure~\ref{fig_5_stirSpec}. The spectra for three different driving levels are roughly the same, with slight deviations at large scales.  The compensated spectra show the typical signatures of numerical turbulence simulations: flatter than the expected power law at low k (appearing upward sloped here), following the expected power law at intermediate k, and steeper than the expected power law at high k as we approach the dissipation scale. We consider that these spectra are sufficiently close to Kolmogorov for our purposes in this study.

\section{Model Prefactors}
\label{app_prefactors}

We list here the normalization co-efficients for the models used in \S\ref{sec_IPMF}.
\begin{enumerate}
\item Simply Tapered Power Law (STPL),
\begin{equation}
c_1 = \frac{1}{M_{\text{min}}^{-\alpha}} \exp{\bigg(\frac{M_\text{min}}{M_{\text{exp}}}\bigg)}.
\end{equation}

\item Variably Tapered Power Law (VTPL),
\begin{equation}
c_2 = \frac{1}{M_{\text{min}}^{-\alpha}} \exp{\bigg[\bigg(\frac{M_\text{min}}{M_{\text{exp}}}\bigg)^{\beta}\bigg]}.
\end{equation}

\item Broken Cumulative Power Law (BCPL),
\begin{align}
c_{31} &= \frac{1}{M_{\text{min}}^{-\alpha}} \\
c_{32} &= \frac{1}{M_{\text{min}}^{-\alpha_1}M_{\text{br}}^{\alpha_1-\alpha_2}}. \nonumber
\end{align}

\item Truncated Power Law (TPL),
\begin{equation}
c_4 = \frac{\alpha}{M_{\text{min}}^{-\alpha} - M_{\text{tr}}^{-\alpha}}.
\end{equation}

\item Broken Power Law (BPL),
\begin{align}
c_{51} &= \frac{1}{M_{\text{min}}^{-\alpha_1}} \bigg[ \frac{1}{\alpha_1} + \bigg( \frac{1}{\alpha_2} - \frac{1}{\alpha_1} \bigg) \bigg( \frac{M_{\text{br}}}{M_{\text{min}}} \bigg)^{-\alpha_1} \bigg]^{-1} \\
c_{52} &=  c_{51} M_{\text{br}}^{\alpha_2-\alpha_1}. \nonumber
\end{align}

\item Three Segment Power Law (TSPL),
\begin{align}
c_{61} &= \frac{1}{M_{\text{min}}^{-\alpha_1}} \bigg[ \frac{1}{\alpha_1} + \bigg( \frac{1}{\alpha_2} - \frac{1}{\alpha_1} \bigg) \bigg( \frac{M_{\text{br},1}}{M_{\text{min}}} \bigg)^{-\alpha_1} + \bigg( \frac{1}{\alpha_3} - \frac{1}{\alpha_2} \bigg) \bigg( \frac{M_{\text{br},1}}{M_{\text{min}}} \bigg)^{\alpha_2-\alpha_1} \bigg( \frac{M_{\text{br},2}}{M_{\text{min}}} \bigg)^{-\alpha_2} \bigg]^{-1} \\
c_{62} &= c_{61} M_{\text{br}}^{\alpha_2 - \alpha_1} \nonumber  \\
c_{63} &= c_{61} M_{\text{br},2}^{\alpha_2 - \alpha_1} M_{\text{br},2}^{\alpha_3 - \alpha_2}. \nonumber
\end{align}
\end{enumerate}

\bibliography{turbulence_refs}

\end{document}